\documentclass[a4paper,11pt]{article}
\usepackage[T1]{fontenc}
\usepackage{lmodern}
\usepackage[margin=25mm]{geometry}
\usepackage{amsmath,amssymb,graphicx,booktabs,tabularx,array,enumitem,xcolor}
\usepackage[numbers,sort&compress]{natbib}
\usepackage{xurl}
\usepackage[section]{placeins}
\usepackage[unicode,colorlinks=true,linkcolor=blue!45!black,
  citecolor=blue!45!black,urlcolor=blue!45!black]{hyperref}
\providecommand{\doi}[1]{doi:\,\href{https://doi.org/#1}{\nolinkurl{#1}}}
\hypersetup{
  pdftitle={Shadows and Images of Black Holes with Purely Nonlinear Scalarization},
  pdfauthor={Hao Yang}}
\graphicspath{{images/}}
\setlist[itemize]{leftmargin=1.6em,itemsep=0.12em,topsep=0.2em}

\newcolumntype{Y}{>{\raggedright\arraybackslash}X}
\newcommand{\bcrit}{b_{\mathrm c}}
\newcommand{\deltash}{\delta_{\mathrm{sh}}}
\newcommand{\rph}{r_{\mathrm{ph}}}
\newcommand{\Omegaph}{\Omega_{\mathrm{ph}}}
\newcommand{\Lambdaph}{\Lambda_{\mathrm{ph}}}
\newcommand{\risco}{r_{\mathrm{ISCO}}}

\newcommand{\Sch}{\mathrm{Sch}}

\title{Shadows and Images of Black Holes with Purely Nonlinear Scalarization}
\author{Hao Yang\thanks{e-mail: \href{mailto:yang@ucas.ac.cn}{\nolinkurl{yang@ucas.ac.cn}}}\\[4pt]
  \small School of Fundamental Physics and Mathematical Sciences, Hangzhou Institute for Advanced Study,\\
  \small UCAS, Hangzhou 310024, China}
\date{}

\begin{document}
\maketitle
\begin{abstract}
We investigate the shadows and radiative images of static, spherically symmetric black holes with purely nonlinear scalarization in Einstein--scalar--Gauss--Bonnet gravity.
The quartic exponential coupling supports scalarized backgrounds without a linear tachyonic instability of Schwarzschild black holes.
We compare mass and coupling dependences and construct images on radially stable backgrounds using optically thin spherical emission and face-on thin disks.
On the sampled radially stable branches, we find smaller shadow radii than those of Schwarzschild black holes at the same mass, with reductions reaching approximately $21.7\%$.
The sampled radially unstable branches show much smaller shadow-size deviations.
The photon-orbit index, which measures radial instability per unit orbital angle, varies nonmonotonically even when the normalized shadow radius changes monotonically.
The index also controls the asymptotic compression of higher-order images, providing information complementary to shadow size.
For spherical emission, radial infall produces a darker center relative to the bright ring than static emission.
For circular disks, we find that smooth metric variations can create an unstable interval separating two stable regions of timelike circular motion.
When the disk occupies the stable region connected to infinity, its inner edge moves outward across this transition, suppressing emission inside the geometric shadow boundary while the normalized shadow radius changes only slightly.
Our results show that purely nonlinear scalarization can alter images through both photon propagation and the orbital structure of emitting matter, with the visible features depending on the emission model.
\end{abstract}

\tableofcontents

\section{Introduction}\label{sec:introduction}
Event-horizon-scale images of M87* and Sagittarius A* obtained by the Event Horizon Telescope provide~\citep{EHT2019M87I,EHT2022SgrAI,EHT2024M87Persistent,EHT2024SgrAPolarization,EHT2025M87Variability} a means of studying photon propagation and emitting matter in the strong gravitational fields of supermassive black holes.
At the geometric level, analytical descriptions of black-hole shadows relate~\citep{PerlickTsupko2022} their boundaries to photon capture and the critical null geodesics of the spacetime.
In an observed image, however, the bright ring and central brightness depression reflect gravitational lensing and photon capture together with the distribution and motion of the emitting plasma.
The geometric shadow boundary, the brightness maximum, and the edge of a dark central region therefore need not coincide~\citep{GrallaHolzWald2019}.
Accordingly, using black-hole images to probe gravity requires understanding how spacetime geometry shapes these features.

A relevant theoretical setting is Einstein--scalar--Gauss--Bonnet gravity, which allows~\citep{AntoniouEtAl2018} black holes to acquire a nontrivial scalar field through a coupling between the scalar and spacetime curvature.
For suitable coupling functions, scalar-free solutions remain solutions of the theory, while sufficiently strong curvature can trigger a tachyonic instability of the scalar perturbation.
The resulting spontaneous scalarization gives rise~\citep{Doneva:2017bvd,SilvaEtAl2018,DonevaEtAl2024Review} to black hole branches with nonzero scalar hair.
%Since the scalar field backreacts on the metric, scalarization can modify the motion of photons and massive particles outside the horizon, allowing geometries with the same ADM mass to have different optical properties~\citep{CunhaHerdeiroRadu2019,FernandesEtAl2024}.
%The existence of scalar hair alone does not ensure radial stability, which depends on the coupling function and the solution branch~\citep{BlazquezSalcedoEtAl2018}.
However, the absence of a linear tachyonic instability does not exclude scalarization.
In particular, for couplings whose leading dependence near the scalar-free state is quartic, the curvature-induced mass term vanishes in the linear scalar equation, but nonlinear interactions can support~\citep{DonevaYazadjiev2022} nontrivial scalar configurations.
Indeed, time evolutions in the decoupling limit show that sufficiently large initial scalar perturbations can settle into nonzero stationary profiles, while smaller perturbations decay.
Beyond the decoupling limit, the corresponding fully coupled equilibrium solutions can have several scalarized branches at the same mass.
Radial perturbation analyses further identify~\citep{BlazquezSalcedoEtAl2022} stable regions of these branches as well as unstable and nonhyperbolic regions.

The optical consequences of scalarization have also been studied, with appreciable changes found~\citep{CunhaHerdeiroRadu2019,FernandesEtAl2024,WangEtAl2024} in black-hole shadows and circular-orbit properties.
For example, studies of spontaneously scalarized black holes and their rotating extensions have shown~\citep{CunhaHerdeiroRadu2019} that, even without rotation, scalarized black holes can have shadow sizes that differ substantially from those of Schwarzschild black holes with the same mass.
In the case of spin-induced scalarization, circular-orbit properties have been found~\citep{FernandesEtAl2024} to differ appreciably from those of Kerr black holes even when the shadow size deviations remain small.
Beyond these geometric properties, images of spontaneously scalarized black holes illuminated by thin disks and static spherical emission demonstrate~\citep{WangEtAl2024} that the optical appearance depends on both the coupling and the emission prescription.
These results therefore motivate extending the comparison to purely nonlinear scalarized branches and following their optical properties across mass, coupling, and radial stability.

More generally, black hole images can be described by three complementary properties: the shadow size, the structure of photon rings, and the brightness distribution.
First, the shadow size measures the extent of the geometric capture region.
Specifically, in the static, spherically symmetric setting, the critical impact parameter $\bcrit$ sets~\citep{PerlickTsupko2022} the capture boundary and the asymptotic geometric shadow radius.
Second, photon-ring structure is characterized by two complementary quantities: the photon-ring width and the photon-orbit index.
The photon-ring width specifies~\citep{GrallaHolzWald2019} the interval of impact parameters occupied by rays with large accumulated orbital angles, and hence the radial extent of the corresponding region in the image plane.
Alongside this width, the photon-orbit index $\chi$ measures~\citep{CardosoEtAl2009,KocherlakotaEtAl2024} the exponential rate at which nearby rays separate radially from the unstable circular photon orbit per unit orbital angle.
The index also controls~\citep{JohnsonEtAl2020,KocherlakotaEtAl2024} the asymptotic concentration of successive higher-order images toward the critical curve.
Finally, the brightness distribution gives the observed intensity at each position in the image and depends on both photon propagation and the spatial distribution and motion of the emitting matter.

Predicting the brightness distribution therefore requires an emission model.
For example, optically thin spherical emission provides~\citep{NarayanJohnsonGammie2019} a simple setting for relating the critical curve to a central brightness depression, while thin-disk images illustrate~\citep{GrallaHolzWald2019} the importance of the emitting region and its multiple images.
However, studies of spherical images have shown~\citep{KocherlakotaRezzolla2022} that degeneracies between geometry and emission can persist, motivating the joint use of several image properties under controlled emission assumptions.
In thin-disk models with circularly orbiting emitters, the spacetime also determines the allowed stable orbits and hence informs the choice of disk inner edge.
In particular, more general black-hole geometries can contain~\citep{DelgadoEtAl2022} disconnected regions of stable timelike circular motion, so the innermost stable orbit and the inner boundary of the stable region accessible from large radius need not coincide.
It is therefore important to assess how different emission models affect the brightness distribution and overall appearance of black-hole images.
%This distinction raises a separate imaging question: how strongly can a change in the adopted disk inner edge affect the brightness distribution when the geometric shadow changes only slightly?

Motivated by these considerations, we investigate the shadows and radiative images of static, spherically symmetric black holes with purely nonlinear scalarization in scalar--Gauss--Bonnet gravity.
Using Schwarzschild black holes as a reference, we examine the mass and coupling dependence of the shadow radius, photon-orbit index, and photon-ring width along the scalarized branches.
Building on this geometric comparison, we then study spherical emission and thin-disk images to determine how emitter motion and the disk inner edge affect the brightness distribution.
Together, the geometric and emission analyses connect the image properties to photon dynamics and the orbital structure of the emitting matter.

The paper is organized as follows.
In Section~\ref{sec:backgrounds}, we introduce the theory, scalarized backgrounds, radial stability, and comparison strategy.
In Section~\ref{sec:geometry_results}, we then present the shadow size, photon-orbit instability, and photon-ring structure together with their mass and coupling dependence.
In Section~\ref{sec:imaging}, we connect these geometric results to spherical emission and thin-disk images, including the role of circular-orbit stability in selecting the disk inner edge.
Finally, in Section~\ref{sec:conclusions}, we summarize the results and discuss future directions.
Throughout the paper, we use geometrized units, $G=c=1$, and the metric signature $(-,+,+,+)$.

\section{Theoretical Model and Black-Hole Backgrounds}\label{sec:backgrounds}
% Preserve former subsection labels as references to the merged section.
\label{subsec:model}
\label{subsec:metric}
\label{subsec:branches}
\label{subsec:comparison_design}
We consider four-dimensional Einstein--scalar--Gauss--Bonnet gravity with a massless real scalar field and no scalar potential. 
The action of the theory is
\begin{equation}
 S=\frac{1}{16\pi}\int d^4x\,\sqrt{-g}\,
 \left[R-2\nabla_\mu\phi\nabla^\mu\phi
       +\lambda^2 f(\phi)\mathcal G\right],
 \label{eq:action}
\end{equation}
where $g=\det(g_{\mu\nu})$, $R$ is the Ricci scalar, $\phi$ is a dimensionless real scalar field, and $\lambda$ is a positive coupling constant with dimensions of length.
The dimensionless function $f(\phi)$ couples the scalar to the curvature through the Gauss--Bonnet invariant,
\begin{equation}
 \mathcal G=R^2-4R_{\mu\nu}R^{\mu\nu}
              +R_{\mu\nu\rho\sigma}R^{\mu\nu\rho\sigma}
 \label{eq:gb_invariant}
\end{equation}
where $R_{\mu\nu}$ and $R_{\mu\nu\rho\sigma}$ are the Ricci and Riemann tensors, respectively.
In this work, we adopt~\citep{DonevaYazadjiev2022} the quartic exponential coupling used to study nonlinear black-hole scalarization,
\begin{equation}
 f(\phi)=\frac{1-e^{-\kappa\phi^4}}{4\kappa},
 \qquad \kappa>0,
 \label{eq:quartic_coupling}
\end{equation}
Here the dimensionless parameter $\kappa$ controls the shape of the coupling function.
There is no background matter or electromagnetic charge. 
%Radiation is subsequently treated as a test field minimally coupled to $g_{\mu\nu}$, without a direct scalar--electromagnetic coupling.

Varying Eq.~\eqref{eq:action} with respect to the inverse metric gives the modified Einstein equations 
%which we write in the form used in Ref.~\citep{HerdeiroEtAl2026},
\begin{equation}
 G_{\mu\nu}+\lambda^2 H_{\mu\nu}=T^\phi_{\mu\nu},
 \label{eq:einstein_equations}
\end{equation}
where $G_{\mu\nu}=R_{\mu\nu}-\tfrac12g_{\mu\nu}R$ is the Einstein tensor.
The tensor $H_{\mu\nu}$ represents the contribution from the scalar--Gauss--Bonnet coupling, while $T^\phi_{\mu\nu}$ is the scalar kinetic source in the normalization of Eq.~\eqref{eq:action}:
\begin{equation}
 \begin{aligned}
 H_{\mu\nu}={}&-R\left(\nabla_\mu\Psi_\nu+\nabla_\nu\Psi_\mu\right)
  -4\nabla^\rho\Psi_\rho G_{\mu\nu}\\
 &+4R_{\mu\rho}\nabla^\rho\Psi_\nu
  +4R_{\nu\rho}\nabla^\rho\Psi_\mu\\
 &-4g_{\mu\nu}R^{\rho\sigma}\nabla_\rho\Psi_\sigma
  +4R^\sigma{}_{\mu\rho\nu}\nabla^\rho\Psi_\sigma,\\
 T^\phi_{\mu\nu}={}&2\nabla_\mu\phi\nabla_\nu\phi
  -g_{\mu\nu}\nabla_\rho\phi\nabla^\rho\phi.
 \end{aligned}
 \label{eq:gb_metric_contribution}
\end{equation}
To express the coupling contribution compactly, we write $f_{,\phi}=df/d\phi$ and introduce its spacetime gradient,
\begin{equation}
 \Psi_\mu=f_{,\phi}(\phi)\nabla_\mu\phi
         =\nabla_\mu f(\phi).
 \label{eq:gb_coupling_gradient}
\end{equation}
%The coefficients in $T^\phi_{\mu\nu}$ follow from the kinetic normalization in Eq.~\eqref{eq:action}.\footnote{The canonically normalized scalar of Ref.~\citep{HerdeiroEtAl2026} is $\varphi=2\phi$. Its parameters obey $\alpha=\lambda^2$ and, for the type~(iii) coupling, $\beta=\kappa$, with $f_{\mathrm{ref}}(\varphi)=f(\varphi/2)$. These identifications give the same field equations.}

Variation of Eq.~\eqref{eq:action} with respect to the scalar field gives
\begin{equation}
 \Box\phi+\frac{\lambda^2}{4}f_{,\phi}(\phi)\mathcal G=0,
 \qquad \Box=\nabla_\mu\nabla^\mu.
 \label{eq:scalar_equation}
\end{equation}
%The normalization of the kinetic term fixes the factor $1/4$ in this equation. 
For the quartic exponential coupling in Eq.~\eqref{eq:quartic_coupling}, the first two scalar derivatives are
\begin{align}
 f_{,\phi}(\phi)&=\phi^3e^{-\kappa\phi^4},\nonumber\\
 f_{,\phi\phi}(\phi)&=(3\phi^2-4\kappa\phi^6)e^{-\kappa\phi^4},
 \label{eq:coupling_derivatives}
\end{align}
so that $f(0)=f_{,\phi}(0)=f_{,\phi\phi}(0)=0$. 
Scalar-free Schwarzschild black holes therefore remain exact solutions.
The linear scalar perturbation $\delta\phi$ about $\phi=0$ satisfies
\begin{equation}
 \Box_{\Sch}\,\delta\phi=0,
 \label{eq:linear_scalar_schwarzschild}
\end{equation}
Here $\Box_{\Sch}$ is the wave operator on the Schwarzschild background, and no curvature-induced tachyonic mass term appears.
In conventional curvature-induced spontaneous scalarization, a coupling with $f_{,\phi\phi}(0)>0$ instead gives~\citep{Doneva:2017bvd,SilvaEtAl2018} the linear scalar perturbation an effective mass squared $m_{\mathrm{eff}}^2=-\lambda^2 f_{,\phi\phi}(0)\mathcal G_{\Sch}/4$.
In a sufficiently strong curvature regime, this term can produce a tachyonic instability, allowing arbitrarily small perturbations with a component along the unstable mode to grow.
Scalarized branches then bifurcate~\citep{Doneva:2017bvd} from the family of Schwarzschild black holes at the onset of this instability.

However, the absence of this linear mechanism does not exclude scalarization.
For the quartic coupling considered here, the scalar--curvature source becomes relevant at finite amplitude.
For suitable black hole masses and coupling parameters, evolutions in the decoupling limit show~\citep{DonevaYazadjiev2022} that small perturbations decay, whereas initial scalar perturbations exceeding a finite threshold can settle into a nonzero stationary profile, leaving a long-lived scalar configuration outside the horizon.
This is the purely nonlinear scalarization mechanism relevant to the present model.
The horizon value $\phi_H\equiv\phi(r_H)$, where $r_H$ is the horizon's areal radius, helps distinguish how the resulting branches connect to scalar-free solutions.
For conventional spontaneous scalarization, $\phi_H$ tends continuously to zero where the scalarized branch meets Schwarzschild black holes.
In the purely nonlinear model, the radially stable scalarized branch does not emerge~\citep{DonevaYazadjiev2022,BlazquezSalcedoEtAl2022} from Schwarzschild black holes through such a linear onset at finite mass, but reaching it entails a finite change in the scalar amplitude.

%It does not determine the stability of the resulting scalarized backgrounds.

%The present study takes $\kappa=100$ and $1000$ and keeps $\lambda$ fixed. 
%We set $\lambda=1$ in the numerical implementation, while reporting dimensional quantities in units of $\lambda$. 
%In particular,
%\begin{equation}
 %\rho=\frac{r}{\lambda},\qquad
 %\mu=\frac{M}{\lambda},\qquad
 %\widehat r_H=\frac{r_H}{\lambda},\qquad
 %\widehat D=\frac{D}{\lambda}
 %\label{eq:dimensionless_background_quantities}
%\end{equation}
%are the background variables, where $M$ and $D$ are defined by the asymptotic expansion below. Optical profiles also use $y=r/M=\rho/\mu$.
%Normalizing a plot by $M$ does not change the value of $M/\lambda$ or make two backgrounds equal in mass.

%Our scalar-charge convention is $\phi=D/r+O(r^{-2})$ at infinity.
%If the scalar is instead written with canonical kinetic normalization, $\phi_{\mathrm{can}}=2\phi$, the kinetic term becomes $-\tfrac12\nabla_\mu\phi_{\mathrm{can}}\nabla^\mu\phi_{\mathrm{can}}$ and the coupling is $f(\phi_{\mathrm{can}}/2)$.
%The corresponding asymptotic coefficient is $Q_s=2D$, which is the charge used in the background workbench.
%We distinguish $D/\lambda$ from $Q_s/\lambda$ whenever background branch diagrams are compared. Since the coupling is even in $\phi$, the transformation $\phi\mapsto-\phi$ leaves the metric unchanged.
%We use positive, nodeless scalar profiles, and their negative counterparts have identical minimally coupled photon observables.

To investigate the optical properties of these scalarized solutions, we restrict attention to static, spherically symmetric, and asymptotically flat backgrounds.
Using the areal radius $r$, for which a sphere has area $4\pi r^2$, we write
\begin{equation}
 ds^2=-A(r)\,dt^2+\frac{dr^2}{B(r)}
       +r^2\bigl(d\theta^2+\sin^2\theta\,d\varphi^2\bigr),
 \qquad \phi=\phi(r),
 \label{eq:background_metric}
\end{equation}
Here $t$ is the static time coordinate, and $(\theta,\varphi)$ are the usual angular coordinates.
The metric functions $A=e^{2\Phi}$ and $B=e^{-2\Lambda}$ specify the time and radial components, with $\Phi$ and $\Lambda$ their logarithmic parametrizations.
The functions $A$ and $B$ are determined independently and are positive in the static exterior. 
In general, $A\ne B$.
The field equations are solved with the full scalar backreaction on the metric. 
In particular, Eq.~\eqref{eq:scalar_equation} reduces to
\begin{equation}
 \phi''+\left(\frac{2}{r}+\frac{A'}{2A}+\frac{B'}{2B}\right)\phi'
       +\frac{\lambda^2}{4B}f_{,\phi}(\phi)\mathcal G=0,
 \label{eq:radial_scalar_equation}
\end{equation}
where a prime denotes differentiation with respect to $r$.
This equation is integrated together with the independent components of the Einstein equations~\eqref{eq:einstein_equations}, evaluated for the ansatz~\eqref{eq:background_metric}.

For a regular nonextremal horizon at $r=r_H$, the leading exterior expansions have the form
\begin{align}
 A(r)&=a_H(r-r_H)+O\bigl((r-r_H)^2\bigr),\nonumber\\
 B(r)&=b_H(r-r_H)+O\bigl((r-r_H)^2\bigr),\nonumber\\
 \phi(r)&=\phi_H+\phi'_H(r-r_H)+O\bigl((r-r_H)^2\bigr),
 \label{eq:horizon_expansion}
\end{align}
Here $a_H=A'(r_H)$ and $b_H=B'(r_H)$ are positive horizon slopes, while $\phi_H$ is finite.
The horizon derivative $\phi'_H$ is not an independent parameter and must be compatible with a regular solution of the coupled field equations.
Substituting Eq.~\eqref{eq:horizon_expansion} into Eqs.~\eqref{eq:einstein_equations} and \eqref{eq:radial_scalar_equation}, and eliminating the metric expansion coefficients, gives the following constraint 
\begin{equation}
 2\lambda^2 r_H^2 f_{,\phi}(\phi_H)(\phi'_H)^2
 +r_H^3\phi'_H+3\lambda^2 f_{,\phi}(\phi_H)=0.
 \label{eq:horizon_slope_constraint}
\end{equation}
For $f_{,\phi}(\phi_H)\ne0$, solving Eq.~\eqref{eq:horizon_slope_constraint} yields~\citep{Doneva:2017bvd} the two algebraic roots
\begin{equation}
 \phi'_{H,\pm}=\frac{r_H}{4\lambda^2 f_{,\phi}(\phi_H)}
                    \left(-1\pm\sqrt{\Delta_H}\right),
 \label{eq:horizon_scalar_roots}
\end{equation}
where the dimensionless discriminant of the horizon constraint is
\begin{equation}
 \Delta_H=1-\frac{24\lambda^4[f_{,\phi}(\phi_H)]^2}{r_H^4}.
 \label{eq:horizon_discriminant}
\end{equation}
Here, we select the plus sign, for which $\phi'_{H,+}\to0$ as $f_{,\phi}(\phi_H)\to0$ at fixed $r_H$ and $\lambda$, and the other root diverges in this limit.
%When the coupling derivative vanishes exactly, Eq.~\eqref{eq:horizon_slope_constraint} likewise gives $\phi'_H=0$.
The selected root can equivalently be written as
\begin{equation}
 \phi'_H=-\frac{6\lambda^2 f_{,\phi}(\phi_H)}
                  {r_H^3(1+\sqrt{\Delta_H})}.
 \label{eq:horizon_scalar_slope}
\end{equation}
We restrict the backgrounds to horizon satisfying
\begin{equation}
 \Delta_H>0
 \quad\Longleftrightarrow\quad
 r_H^4>24\lambda^4[f_{,\phi}(\phi_H)]^2.
 \label{eq:horizon_regularity}
\end{equation}
The limiting case $\Delta_H=0$, at which the two local roots coalesce, is excluded because the quadratic constraint alone does not establish regularity of the higher-order horizon expansion there.
The radial metric slope $b_H=B'(r_H)$ follows from the $rr$ component of the Einstein equations~\eqref{eq:einstein_equations} evaluated at the horizon
\begin{equation}
 b_H=\frac{1}{r_H+2\lambda^2 f_{,\phi}(\phi_H)\phi'_H}.
 \label{eq:horizon_metric_slope}
\end{equation}
The normalization of $a_H$ is fixed by imposing $A(\infty)=1$.
These local conditions supply horizon data, but do not by themselves establish a globally regular or dynamically admissible exterior.

At spatial infinity we impose
\begin{equation}
 A(r)\longrightarrow1,\qquad B(r)\longrightarrow1,\qquad
 \phi(r)\longrightarrow0.
 \label{eq:asymptotic_boundary_conditions}
\end{equation}
The leading expansions then define the Arnowitt--Deser--Misner (ADM) mass $M$ and scalar charge $D$:
\begin{align}
 A(r)&=1-\frac{2M}{r}+O(r^{-2}),\nonumber\\
 B(r)&=1-\frac{2M}{r}+O(r^{-2}),\nonumber\\
 \phi(r)&=\frac{D}{r}+O(r^{-2}).
 \label{eq:asymptotic_expansions}
\end{align}
Both $M$ and $D$ have dimensions of length in the adopted units, so $M/\lambda$ and $D/\lambda$ are their dimensionless values.
%The scalar charge is the coefficient of the asymptotic scalar tail, not an electromagnetic charge.
The scalar boundary value is fixed rather than treated as an additional parameter.

Together, the horizon and asymptotic conditions define the boundary-value problem for the scalarized backgrounds.
We adjust the horizon data to satisfy $\phi(\infty)=0$ and follow the resulting solution branches, extracting $M$ and $D$ from Eq.~\eqref{eq:asymptotic_expansions}.

Since the scalar-free solution remains available in the same theory, we use a Schwarzschild black hole with the same ADM mass as the reference geometry,
\begin{equation}
 A_{\Sch}(r)=B_{\Sch}(r)=1-\frac{2M}{r},\qquad
 \phi_{\Sch}=0,\qquad r_{H,\Sch}=2M.
 \label{eq:schwarzschild_background}
\end{equation}
The scalarized backgrounds need not satisfy $r_H=2M$.
Consequently, pointwise comparisons of static exterior fields are restricted to $r>\max(r_H,2M)$.
Matching the horizon radius would not, in general, match the ADM mass.

\begin{figure}[htbp]
 \centering
 \includegraphics[width=\linewidth]{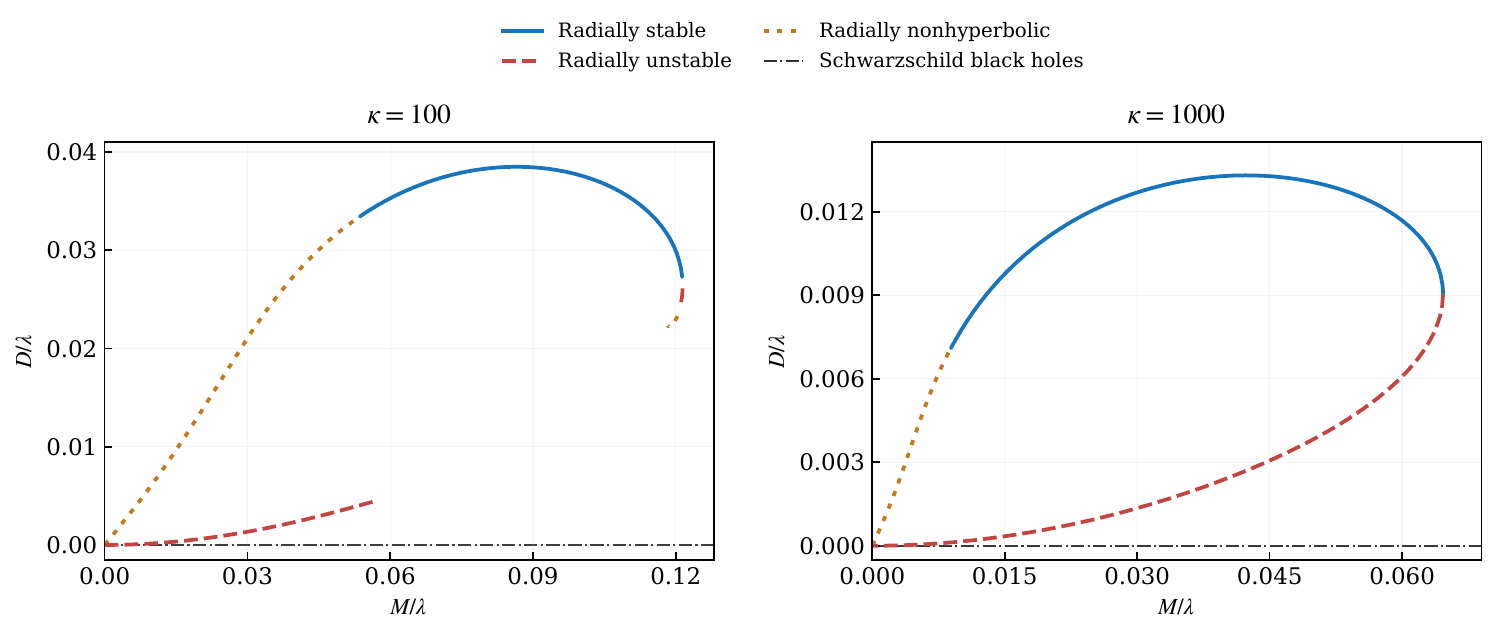}
 \caption{The dimensionless scalar charge $D/\lambda$ is shown as a function of the dimensionless ADM mass $M/\lambda$ for scalarized solutions at $\kappa=100$ (left) and $\kappa=1000$ (right).
 Blue solid curves denote radially stable configurations, red dashed curves denote radially unstable configurations, and orange dotted curves denote radial loss of hyperbolicity.
 The black dash-dotted line represents Schwarzschild black holes. 
 }
 \label{fig:branches}
\end{figure}

In addition to the scalar-free solution, different scalarized configurations can coexist~\citep{DonevaYazadjiev2022} at the same mass.
To identify these solution families and their stability properties, we present the scalar-charge--mass diagrams in Fig.~\ref{fig:branches} at fixed coupling scale $\lambda$.
The left and right panels show the representative couplings $\kappa=100$ and $1000$, respectively, with line styles distinguishing radial stability and loss of hyperbolicity.
The axes are the dimensionless ADM mass $M/\lambda$ and scalar charge $D/\lambda$, with $D$ defined by Eq.~\eqref{eq:asymptotic_expansions}.
Only positive-scalar solutions are needed because the symmetry $\phi\mapsto-\phi$ leaves the metric unchanged.
%Thus the mass alone does not specify the background: the scalar charge and the location along the solution family also distinguish the geometries whose optical properties will be compared.

Regularity of a static solution does not by itself establish its dynamical stability.
We therefore classify the branches in Fig.~\ref{fig:branches} using radial perturbations of the corresponding backgrounds.
Eliminating the metric perturbations in the spherically symmetric sector gives~\citep{BlazquezSalcedoEtAl2022} a scalar perturbation equation of the form
\begin{equation}
 \mathcal K(r)\,\partial_t^2\delta\phi
 -\partial_r^2\delta\phi+C_1(r)\,\partial_r\delta\phi
 +U(r)\,\delta\phi=0,
 \label{eq:radial_perturbation_master}
\end{equation}
Here $\delta\phi$ is the linear scalar perturbation, $\mathcal K$ multiplies its second time derivative, $C_1$ multiplies its first radial derivative, and $U$ is the effective radial perturbation potential.
All three coefficients are fixed by the static background.
The highest-derivative terms describe a wave equation with real radial characteristic speeds when $\mathcal K>0$.
A zero or sign change of $\mathcal K$ therefore signals a loss of radial hyperbolicity, meaning that this wave-propagation description breaks down.
On a hyperbolic background, a mode $\delta\phi=e^{\sigma t}u_{\mathrm{rad}}(r)$ with growth rate $\sigma>0$ and radial profile $u_{\mathrm{rad}}$ satisfying the black-hole boundary conditions represents a radial instability.
We refer to configurations as radially stable when the available finite-domain analysis finds a positive $\mathcal K$ and no growing mode, and as radially unstable when a growing mode is detected on a hyperbolic background.
This classification is restricted to the available linear radial analysis and does not establish stability against arbitrary perturbations.

For $\kappa=1000$ in the right panel, we find two scalar-charge branches at the same black-hole mass, with the displayed sequence reaching a maximum mass of $M/\lambda\simeq0.06463$.
The lower-charge branch is radially unstable, whereas the upper-charge branch contains a radially stable segment.
Following the upper-charge branch toward smaller masses leads to the orange nonhyperbolic region.
The branch structure is more involved at $\kappa=100$, shown in the left panel.
A separate low-charge radially unstable branch extends to $M/\lambda\simeq0.05688$.
The upper-charge branch contains a radially stable segment and turns near $M/\lambda\simeq0.1214$, where a short radially unstable continuation is also present.
The low-mass end of the upper-charge branch is nonhyperbolic.

The coexistence of branches with different scalar charges and radial stability motivates three comparisons that separate mass, coupling, and branch effects on the optical properties.
In all three comparisons, we keep $\lambda$ fixed and characterize each scalarized background by $\kappa$, $M/\lambda$, and its location on the solution family.
Schwarzschild black holes with the same ADM mass provide the common reference through Eq.~\eqref{eq:schwarzschild_background}.

The first comparison follows the mass dependence along each branch at fixed $\kappa$. 
The analysis covers the radially stable segments over approximately
\begin{equation}
 \begin{aligned}
  \kappa=100:&\qquad 0.05363\lesssim M/\lambda\lesssim0.12135,\\
  \kappa=1000:&\qquad 0.009219\lesssim M/\lambda\lesssim0.06457.
 \end{aligned}
 \label{eq:stable_mass_coverage}
\end{equation}
We follow the radially unstable segments separately.
Nonhyperbolic configurations are excluded from the optical comparisons.

The second comparison holds the ADM mass fixed while varying $\kappa$ over the ten values $100,200,\ldots,1000$.
With $\lambda$ unchanged, this varies the coupling function at the same mass scale and tests how the scalarized geometry responds.
The common interval of the ten radially stable segments is approximately $0.05363\lesssim M/\lambda\lesssim0.06457$, with its lower and upper bounds set by the $\kappa=100$ and $1000$ segments in Eq.~\eqref{eq:stable_mass_coverage}, respectively.

The third comparison fixes both $\kappa$ and $M/\lambda$, and then contrasts radially stable and radially unstable configurations with Schwarzschild black holes. 
The masses $M/\lambda=0.05365$ and $0.05550$ allow this comparison at $\kappa=100$ and $1000$.
Comparing these coexisting backgrounds tests whether optical similarity to Schwarzschild black holes is accompanied by similar radial stability.

\section{Shadow Size, Photon-Orbit Instability, and Photon-Ring Structure}
\label{sec:methods}
\label{sec:geometry_results}
We examine the geometric quantities that underlie the radiative images in Section~\ref{sec:imaging}, presenting their theoretical definitions together with their behavior on the scalarized backgrounds.
Starting from null geodesics, we determine the shadow size and compare its mass and coupling dependence.
We then introduce the photon-orbit index and relate its variation to near-critical photon propagation, before examining the number of orbits and photon-ring width.
The comparisons follow branches at fixed $\kappa$ and use matched ADM masses across couplings, with $\lambda$ held fixed.

\subsection{Null geodesics and the optical potential}
\label{subsec:null_geodesics}
We treat light in the geometric-optics approximation, with photons minimally coupled to the metric~\eqref{eq:background_metric} and with negligible backreaction. 
In the absence of a direct scalar--photon coupling, the scalar field affects photon propagation only through the background geometry. 
Spherical symmetry allows each geodesic to be placed in the equatorial plane, $\theta=\pi/2$. 
Let a dot denote differentiation with respect to an affine parameter $\eta$.
The time-translation and rotational symmetries give the conserved photon energy $E$ and angular momentum $L$,
\begin{equation}
 E=A(r)\dot t,\qquad L=r^2\dot\varphi,\qquad b=\frac{L}{E}.
 \label{eq:null_constants}
\end{equation}
%where $E>0$ for future-directed rays in the static exterior. 
%We choose the angular orientation so that $L\geq0$. 
Their ratio defines the impact parameter $b$, which has dimensions of length and is unchanged by a rescaling of the affine parameter.
With $A(\infty)=1$, $b$ specifies the transverse offset of an incoming ray at infinity.
We denote the accumulated orbital angle along a ray by $\psi$, with $d\psi=|d\varphi|$ in the chosen orbital plane and zero set at the starting point.

Substituting Eq.~\eqref{eq:null_constants} into the null constraint $-A\dot t^2+\dot r^2/B+r^2\dot\varphi^2=0$ gives the radial equation of motion,
\begin{equation}
 \dot r^2=\frac{B(r)}{A(r)}E^2
 \left[1-b^2\frac{A(r)}{r^2}\right].
 \label{eq:null_radial_motion}
\end{equation}
The factor $A/r^2$ determines the allowed radial region and therefore motivates the optical potential,
\begin{equation}
 U_{\mathrm{opt}}(r)\equiv\frac{A(r)}{r^2},\qquad
 \frac{A(r)}{B(r)}\dot r^2+L^2U_{\mathrm{opt}}(r)=E^2.
 \label{eq:optical_potential}
\end{equation}
Since $A,B>0$ outside the horizon, a nonradial ray can access only regions where $U_{\mathrm{opt}}\leq b^{-2}$. 
Therefore, a simple root of $U_{\mathrm{opt}}(r)=b^{-2}$ is a radial turning point. 
Thus the potential determines where a ray can propagate, while the factor $B/A$ determines its radial motion along that allowed region.

A circular null orbit at $r=r_0>r_H$ requires both vanishing radial velocity and vanishing radial acceleration. 
Applying these conditions to the radial geodesic equation yields~\citep{CardosoEtAl2009}
\begin{equation}
 b^2U_{\mathrm{opt}}(r_0)=1,\qquad
 U_{\mathrm{opt}}'(r_0)=0,
 \label{eq:circular_null_conditions}
\end{equation}
where primes denote derivatives with respect to $r$. 
In particular,
\begin{equation}
 U_{\mathrm{opt}}'(r)=\frac{rA'(r)-2A(r)}{r^3}
 \quad\Longrightarrow\quad r_0A'(r_0)-2A(r_0)=0.
 \label{eq:photon_orbit_condition}
\end{equation}
An exterior maximum, $U_{\mathrm{opt}}''(r_0)<0$, corresponds to an unstable circular photon orbit, whereas a minimum corresponds to a stable one.
Rotating the orbital plane generates the associated photon sphere.
%This stability concerns photon motion on a fixed geometry and is distinct from the radial stability of the black-hole background discussed in Section~\ref{subsec:branches}. 
%Although the circular-orbit radius is fixed by $A$, the rate of departure from an unstable orbit also depends on $B$, as discussed in Section~\ref{subsec:chi_definition}.

For the asymptotically flat backgrounds considered here, $U_{\mathrm{opt}}$ vanishes at the horizon and tends to zero at infinity.
For a single nondegenerate exterior maximum, an incoming ray with $b^{-2}$ above the maximum crosses the potential barrier and is captured.
If $b^{-2}$ lies below the maximum, the ray instead reaches an outer turning point and returns to infinity.
If several maxima are present, the capture threshold is set by the largest barrier along the path from infinity to the horizon, rather than necessarily by the outermost circular orbit. 
We denote the radius of the unstable orbit controlling this threshold by $\rph$. 
Its relation to the apparent shadow size is developed in Section~\ref{subsec:shadow_size}.

\subsection{Shadow size and its parameter dependence}
\label{subsec:shadow_size}
The capture threshold identified in Section~\ref{subsec:null_geodesics} defines the geometric shadow boundary for a distant observer.
The critical impact parameter $\bcrit$ is fixed by the unstable orbit controlling capture.
Therefore, the first condition in Eq.~\eqref{eq:circular_null_conditions} gives
\begin{equation}
 \bcrit=\frac{1}{\sqrt{U_{\mathrm{opt}}(\rph)}}
       =\frac{\rph}{\sqrt{A(\rph)}},\qquad
 \bcrit^{-2}=\max_{r>r_H}U_{\mathrm{opt}}(r).
 \label{eq:critical_impact_parameter}
\end{equation}
%To compare backgrounds with different ADM masses, we use the dimensionless shadow radius $\bcrit/M$. 
For a Schwarzschild black hole of the same mass, $\rph=3M$ and $A(\rph)=1/3$, giving $\bcrit^{\Sch}=3\sqrt{3}M$.
We therefore define the fractional shadow-size deviation as
\begin{equation}
 \deltash\equiv\frac{\bcrit-\bcrit^{\Sch}}{\bcrit^{\Sch}}
 =\frac{\bcrit/M}{3\sqrt{3}}-1.
 \label{eq:shadow_deviation}
\end{equation}
Positive and negative values indicate a larger and smaller geometric shadow, respectively, relative to a Schwarzschild black hole at the same ADM mass.

These definitions characterize the capture geometry. 
In an image produced by nearby emitting matter, both the brightness maximum and the extent of the central dark region depend~\citep{GrallaHolzWald2019} on the emission distribution, gravitational redshift, and radiative transfer, and need not coincide with the critical curve. 
In particular, rays traced toward the horizon can encounter emitting material before reaching it, so the region inside the geometric shadow need not be completely dark. 
We use $\bcrit/M$ and $\deltash$ to compare the geometries, and examine their relation to the brightness profiles in Section~\ref{sec:imaging}.

\subsubsection{Mass dependence along scalarized branches at fixed \texorpdfstring{$\kappa$}{kappa}}
\label{subsec:fixed_kappa}
\begin{figure}[htbp]
 \centering
 \includegraphics[width=\linewidth]{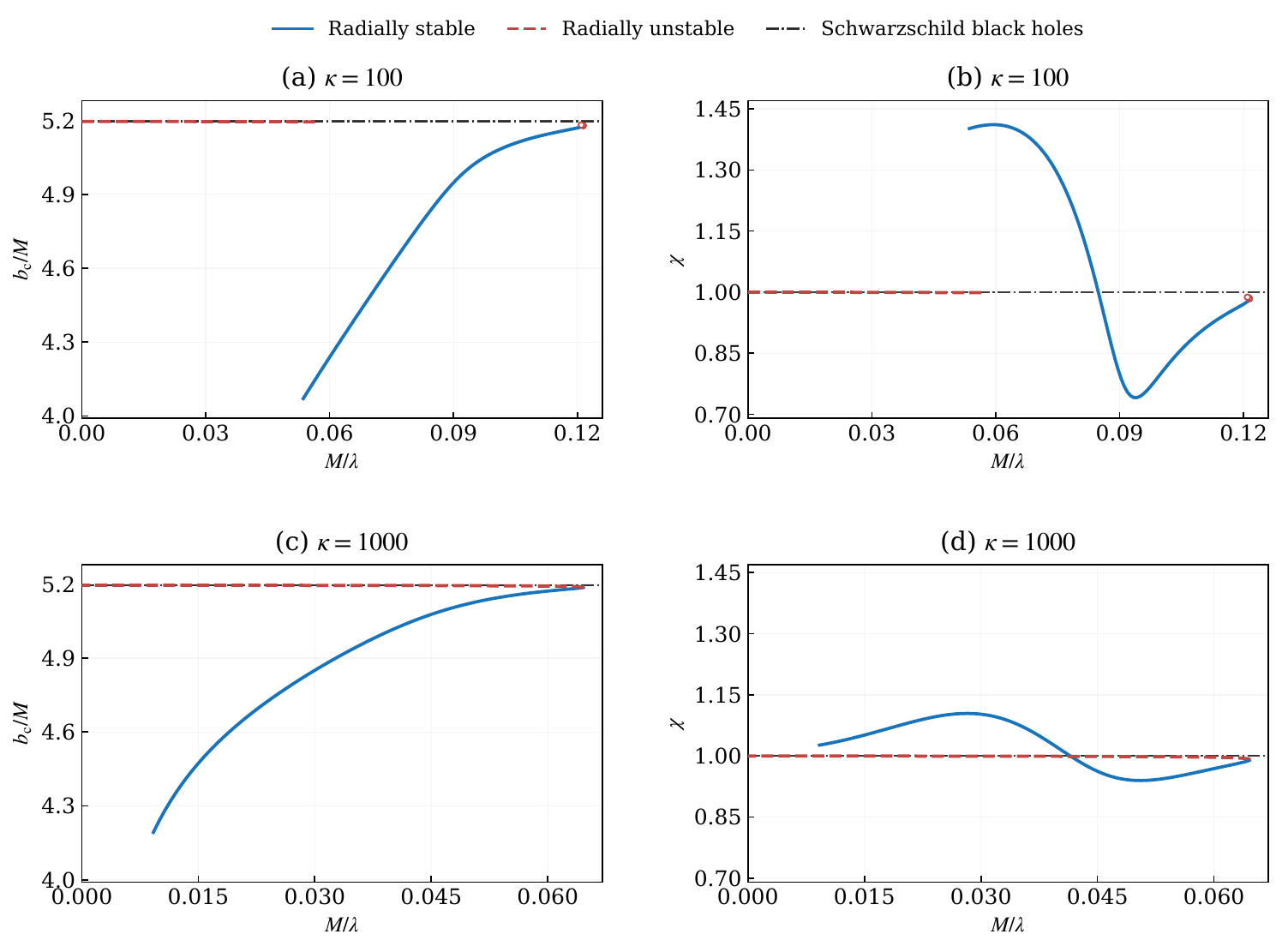}
 \caption{The dimensionless shadow radius $\bcrit/M$ (left column) and photon-orbit index $\chi$ (right column) are shown as functions of mass at fixed $\lambda$, for $\kappa=100$ (top row) and $\kappa=1000$ (bottom row).
 Blue solid and red dashed curves denote radially stable and radially unstable backgrounds, respectively.
 Open circles identify the short unstable continuation near the high-mass turning point for $\kappa=100$.
 Black dash-dotted lines give the values $3\sqrt{3}$ and $1$ for Schwarzschild black holes.
 Each column uses a common vertical scale, while the rows cover their respective mass intervals.
 Curves connect the evaluated backgrounds in branch order, with distinct segments kept separate.}
 \label{fig:fixed_kappa}
\end{figure}

To determine how the shadow changes along each scalarized solution family, we compare the mass dependence at fixed $\kappa=100$ and $1000$ in Fig.~\ref{fig:fixed_kappa}, keeping $\lambda$ fixed.
The upper and lower rows cover the respective mass ranges of the two couplings and distinguish radially stable and unstable backgrounds, with Schwarzschild black holes as the reference.
The left column gives the normalized shadow radius $\bcrit/M$, discussed here, while the right column gives the photon-orbit index $\chi$, whose definition and results follow in Section~\ref{subsec:chi_definition}.

Along both radially stable branches, we find that $\bcrit/M$ increases with mass while remaining below $3\sqrt{3}$.
For $\kappa=100$, the normalized radius increases from approximately $4.0705$ at $M/\lambda\simeq0.05363$ to $5.1763$ at $M/\lambda\simeq0.12135$, corresponding to a change from $\deltash\simeq-21.66\%$ to $-0.381\%$.
For $\kappa=1000$, the normalized radius increases from $4.1926$ at $M/\lambda\simeq0.009219$ to $5.1867$ at $M/\lambda\simeq0.06457$, with $\deltash$ changing from approximately $-19.31\%$ to $-0.181\%$.
Thus, for either coupling, the shadow is smaller than that of a Schwarzschild black hole of the same mass, and the largest reduction occurs at the low-mass end of the stable segment.
By comparison, the radially unstable segments in Fig.~\ref{fig:fixed_kappa} remain close to the reference values for Schwarzschild black holes in both $\bcrit/M$ and $\chi$.
Their optical differences are much less apparent than the pronounced variations found along the radially stable branches.
Within the sampled solution families, the stable branches therefore provide the clearest departures in shadow size and photon-orbit index.
%In particular, the $\kappa=1000$ backgrounds are not uniformly close to Schwarzschild black holes: their high-mass shadows have nearly the same sizes as those of Schwarzschild black holes, but their low-mass shadows can be substantially smaller.

\begin{figure}[htbp]
 \centering
 \includegraphics[width=\linewidth]{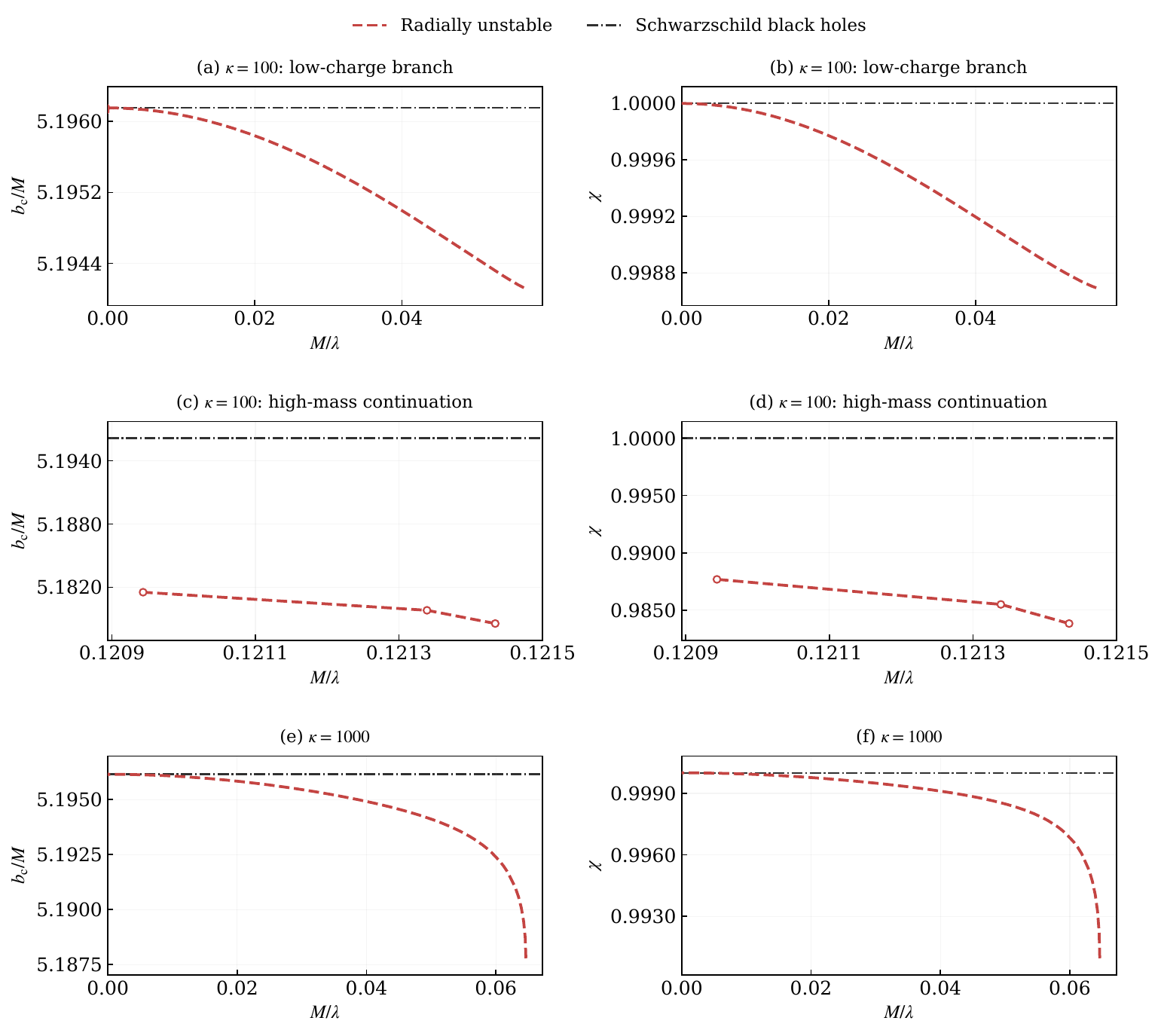}
 \caption{The radially unstable segments in Fig.~\ref{fig:fixed_kappa} are shown separately through $\bcrit/M$ (left) and $\chi$ (right).
 From top to bottom, the rows correspond to the low-charge red dashed branch in the left panel of Fig.~\ref{fig:branches} ($\kappa=100$), the short red dashed continuation near the rightmost turning point of that panel ($\kappa=100$), and the lower-charge red dashed branch in the right panel ($\kappa=1000$).
  Open circles mark the three calculated backgrounds on the short continuation.
 Black dash-dotted lines denote Schwarzschild black holes.}
 \label{fig:unstable_optics}
\end{figure}

To quantify the small departures that are difficult to distinguish on the scales of Fig.~\ref{fig:fixed_kappa}, we display the same unstable backgrounds separately in Fig.~\ref{fig:unstable_optics}, with a row for each connected segment.
The left panels give $\bcrit/M$, while the right panels give $\chi$, discussed in Section~\ref{subsec:chi_fixed_kappa}.
All three rows concern hyperbolic backgrounds with a growing radial mode, and the orange nonhyperbolic segments in Fig.~\ref{fig:branches} are excluded.

The upper row of Fig.~\ref{fig:unstable_optics} corresponds to the low-charge red dashed curve near the bottom of the left panel of Fig.~\ref{fig:branches}, for $\kappa=100$.
At masses where the low-charge branch coexists with the upper branch, the two backgrounds have different nonzero scalar charges, whereas Schwarzschild black holes lie at $D=0$.
The computed low-charge branch extends to $M/\lambda\simeq0.05688$, and its shadow-size deviation satisfies $|\deltash|\lesssim0.04\%$.

The middle row of Fig.~\ref{fig:unstable_optics} corresponds to the short red dashed segment at the rightmost bend of the $\kappa=100$ sequence in the left panel of Fig.~\ref{fig:branches}.
The short segment continues the main solution sequence containing the blue radially stable branch.
Near the maximum mass, the sequence develops a radially unstable portion and bends back toward smaller masses and scalar charges before entering the orange nonhyperbolic region.
The high-mass unstable continuation is separate from the low-charge branch and spans approximately $0.12094\lesssim M/\lambda\lesssim0.12143$, with $\deltash\simeq-(0.28\text{--}0.34)\%$.

The lower row of Fig.~\ref{fig:unstable_optics} corresponds to the lower-charge red dashed branch in the right panel of Fig.~\ref{fig:branches}, for $\kappa=1000$.
The upper- and lower-charge branches form the two sides of a single sequence that turns near $M/\lambda\simeq0.06463$, with the red branch representing the radially unstable side.
Where the red and blue branches coexist at the same mass, the two configurations have different scalar charges and radial stability.
Over the displayed unstable branch, the maximum shadow-size reduction is approximately $0.161\%$.

Across the displayed unstable segments, the shadow-size deviation from Schwarzschild black holes does not exceed approximately $0.34\%$, well below the reductions of about $20\%$ reached at the low-mass ends of the stable branches.
At the representative matched masses $M/\lambda=0.05365$ and $0.05550$, the radially stable solutions also have smaller shadows than their unstable counterparts for both couplings.

The full-branch sequences described above span different mass intervals for the two couplings.
To isolate the coupling dependence, we next compare the shadow sizes at matched ADM masses.

\subsubsection{Coupling dependence at matched ADM mass}\label{subsec:matched_mass}
\begin{figure}[htbp]
 \centering
 \includegraphics[width=\linewidth]{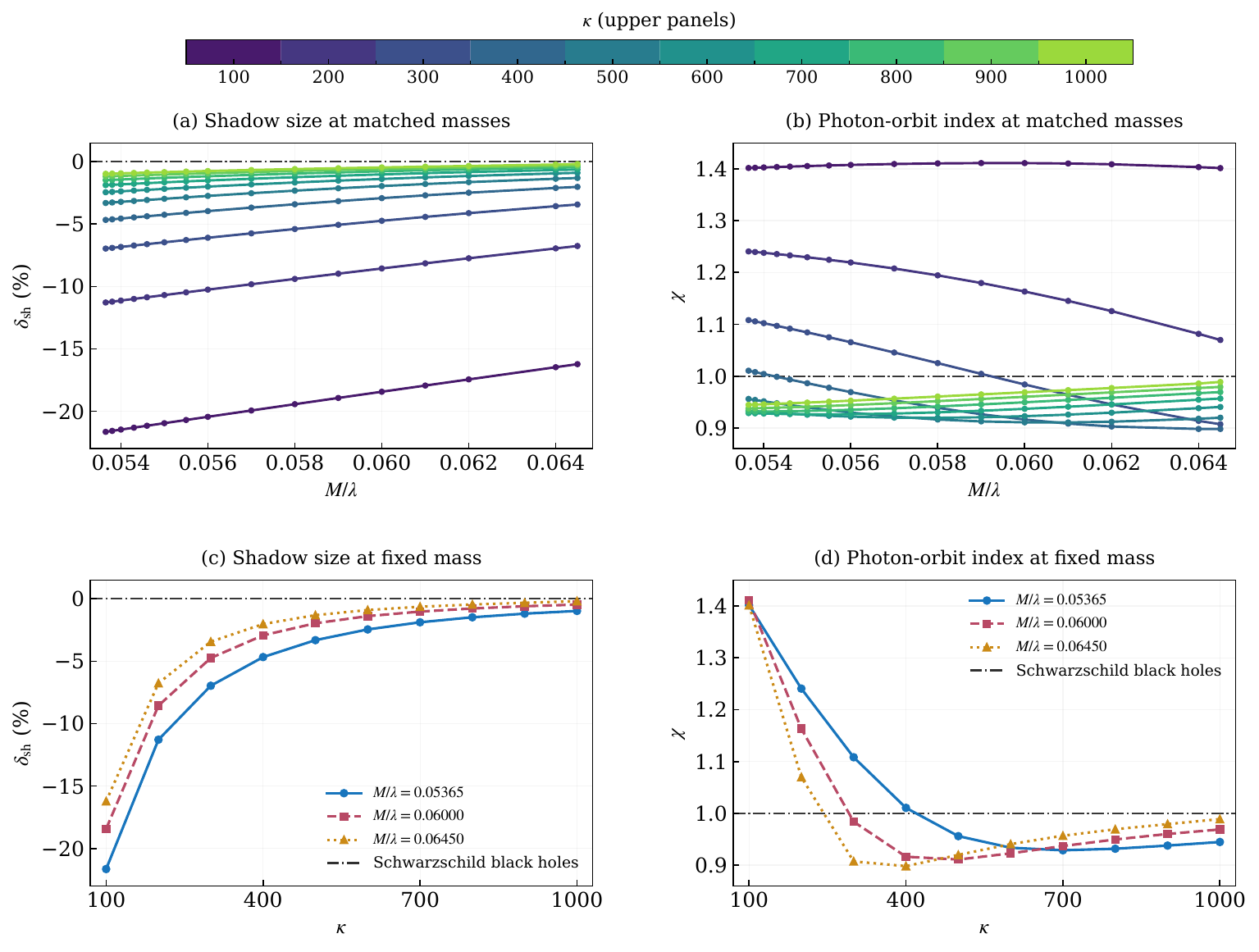}
 \caption{The shadow-size deviation $\deltash$, displayed as a percentage (left column), and photon-orbit index $\chi$ (right column) are compared for radially stable scalarized black holes at fixed $\lambda$.
  The upper panels show sixteen matched masses for each of the ten couplings $\kappa=100,200,\ldots,1000$, distinguished by the color bar.
  The lower panels show the coupling dependence at $M/\lambda=0.05365$, $0.06000$, and $0.06450$, using the same backgrounds.
 Black dash-dotted lines give the reference values for Schwarzschild black holes.}
 \label{fig:matched_mass}
\end{figure}

To separate the effect of the coupling from differences in black-hole mass, we compare radially stable solutions at matched ADM masses in Fig.~\ref{fig:matched_mass}, keeping $\lambda$ fixed.
The ten couplings $\kappa=100,200,\ldots,1000$ share the mass interval $0.05363\lesssim M/\lambda\lesssim0.06457$.
The figure uses sixteen matched masses spanning $0.05365\leq M/\lambda\leq0.06450$, with denser coverage near the lower end, for a total of 160 backgrounds.
The upper panels follow the mass dependence at each coupling, while the lower panels compare the couplings at $M/\lambda=0.05365$, $0.06000$, and $0.06450$.
The left column shows the shadow-size deviation $\deltash$ relative to Schwarzschild black holes, and the right column shows the photon-orbit index discussed in Section~\ref{subsec:chi_matched_mass}.

From the left panels, we find that the shadow radius increases monotonically with $\kappa$ at every matched mass, while remaining below the value for Schwarzschild black holes.
For example, at $M/\lambda=0.06000$, the size deviation is approximately $-18.44\%$, $-1.956\%$, and $-0.460\%$ for $\kappa=100$, $500$, and $1000$, respectively.
At each fixed coupling, the shadow also grows relative to that of Schwarzschild black holes as the mass increases across this common interval.
The difference between the two endpoint couplings narrows with increasing mass: the $\kappa=1000$ shadow radius exceeds the $\kappa=100$ radius by approximately $26.4\%$ at $M/\lambda=0.05365$ and $19.1\%$ at $M/\lambda=0.06450$.

\FloatBarrier

By Eq.~\eqref{eq:critical_impact_parameter}, the increase of $\bcrit/M$ corresponds to a decrease of the dimensionless capture-barrier height $M^2U_{\mathrm{opt}}(\rph)$.
The local instability of rays near this barrier provides additional information, which we now examine through the photon-orbit index.

\subsection{Photon-orbit instability and near-critical behavior}
\label{subsec:chi_definition}
\label{subsec:optical_interpretation}
The instability of the circular photon orbit determines how rapidly nearby rays depart from it. 
%We measure this rate using the asymptotically normalized coordinate time $t$, with $A(\infty)=1$.
Dividing Eq.~\eqref{eq:null_radial_motion} by $\dot t^{\,2}=E^2/A^2$ gives
\begin{equation}
 \left(\frac{dr}{dt}\right)^2
 =A(r)B(r)\left[1-b^2U_{\mathrm{opt}}(r)\right].
 \label{eq:null_coordinate_time}
\end{equation}
Consider a nondegenerate unstable orbit with $U_{\mathrm{opt}}''(\rph)<0$, and write $r=\rph+\delta r$, where $\delta r$ is a small radial displacement.
At $b=\bcrit$, the constant and linear terms in the bracket vanish.
The quadratic expansion gives the growing and decaying radial modes
\begin{equation}
 \delta r(t)\propto e^{\pm\Lambdaph t},\qquad
 \Lambdaph^2=-\frac{B(\rph)\rph^2}{2}
 U_{\mathrm{opt}}''(\rph).
 \label{eq:photon_lyapunov_exponent}
\end{equation}
Here $\Lambdaph>0$ denotes~\citep{CardosoEtAl2009} the coordinate-time Lyapunov exponent, with dimensions of inverse length.

The orbital angular frequency, defined as a positive magnitude, is
\begin{equation}
 \Omegaph\equiv\left|\frac{d\varphi}{dt}\right|_{r=\rph}
 =\frac{\sqrt{A(\rph)}}{\rph}=\frac{1}{\bcrit}.
 \label{eq:photon_orbital_frequency}
\end{equation}
We define the dimensionless photon-orbit index by
\begin{equation}
 \chi\equiv\frac{\Lambdaph}{\Omegaph}
 =\bcrit\Lambdaph
 =\left[\frac{B(r)}{2A(r)}
 \left(2A(r)-r^2A''(r)\right)\right]_{r=\rph}^{1/2},
 \label{eq:photon_orbit_index}
\end{equation}
where the last equality uses $\rph A'(\rph)=2A(\rph)$.
Since the accumulated orbital angle satisfies $\Delta\psi\simeq\Omegaph\Delta t$ near the circular orbit, the growing mode behaves as $|\delta r|\propto e^{\chi\Delta\psi}$.
Thus $\chi$ measures radial instability per unit orbital angle, and larger $\chi$ means faster separation over the same angular interval.

The same index governs the accumulation of orbital angle near the capture threshold. 
For scattering rays with $b>\bcrit$ and with the source and observer outside the photon sphere, the total accumulated angle, including both incoming and outgoing segments, has the leading behavior
\begin{equation}
 \Delta\psi=-\frac{1}{\chi}\ln\epsilon+C+o(1),\qquad
 \epsilon\equiv\frac{b}{\bcrit}-1\longrightarrow0^+.
 \label{eq:near_critical_angle}
\end{equation}
Equation~\eqref{eq:near_critical_angle} gives~\citep{Bozza2002,KocherlakotaEtAl2024} the strong-deflection expansion for a single nondegenerate unstable photon orbit.
Here $\epsilon$ is the fractional distance from the critical impact parameter, and $o(1)$ denotes terms that vanish as $\epsilon\to0^+$.
The finite constant $C$ depends on the geometry and the endpoints, whereas the logarithmic coefficient is fixed by $\chi$.
The derivation and its relation to asymptotic image widths are given in Appendix~\ref{app:photon_scaling}.
A smaller $\chi$ therefore gives a stronger logarithmic increase in the accumulated orbital angle as the critical impact parameter is approached.

For a fixed emitting radial interval, consider successive images labeled by order $n$ such that each additional order adds an orbital angle of $\pi$.
If $w_n$ denotes the width of the corresponding image interval in impact-parameter space, Eq.~\eqref{eq:near_critical_angle} implies
\begin{equation}
 \lim_{n\to\infty}\frac{w_{n+1}}{w_n}=e^{-\pi\chi},\qquad
 \gamma_{\mathrm{half}}\equiv\pi\chi.
 \label{eq:half_orbit_compression}
\end{equation}
The half-orbit lensing exponent $\gamma_{\mathrm{half}}$ is the logarithmic compression per additional half orbit, following the convention used in Ref.~\citep{KocherlakotaEtAl2024}.
Larger $\chi$ produces stronger compression between successive high-order bands. 
%This asymptotic geometric relation does not by itself determine their brightness or flux, and finite-order widths can differ from it.
%The ring sequences and their source geometry are discussed in Section~\ref{sec:rays}.

The shadow size and the photon-orbit index probe different properties of the optical barrier. 
Equations~\eqref{eq:critical_impact_parameter} and~\eqref{eq:photon_lyapunov_exponent} imply
\begin{equation}
 \left(\frac{\bcrit}{M}\right)^{-2}
 =M^2U_{\mathrm{opt}}(\rph),\qquad
 \chi^2=-\frac{B(\rph)\rph^2}{2}
 \frac{U_{\mathrm{opt}}''(\rph)}{U_{\mathrm{opt}}(\rph)}.
 \label{eq:optical_height_and_curvature}
\end{equation}
The dimensionless barrier height fixes $\bcrit/M$: a higher peak gives a smaller shadow. 
The index $\chi$ additionally depends on the curvature of the optical potential at its maximum, the photon-sphere radius, and the radial metric function $B$.
Consequently, backgrounds with equal normalized shadow sizes can have~\citep{KocherlakotaEtAl2024} different near-critical photon dynamics.

For the metric of Schwarzschild black holes in Eq.~\eqref{eq:schwarzschild_background}, the optical potential and its first derivative are
\begin{equation}
 U_{\mathrm{opt}}^{\Sch}(r)=\frac{1}{r^2}-\frac{2M}{r^3},\qquad
 \frac{dU_{\mathrm{opt}}^{\Sch}}{dr}
 =-\frac{2(r-3M)}{r^4}.
 \label{eq:schwarzschild_optical_potential}
\end{equation}
The unique exterior extremum is at $r=3M$, where $U_{\mathrm{opt}}^{\Sch}=1/(27M^2)$ and $d^2U_{\mathrm{opt}}^{\Sch}/dr^2=-2/(81M^4)<0$.
Substitution into the expressions above gives the reference values
\begin{equation}
 \begin{aligned}
  \frac{\rph^{\Sch}}{M}&=3,&
  \frac{\bcrit^{\Sch}}{M}&=3\sqrt{3},&
  \deltash^{\Sch}&=0,\\
  M\Omegaph^{\Sch}&=\frac{1}{3\sqrt{3}},&
  M\Lambdaph^{\Sch}&=\frac{1}{3\sqrt{3}},&
  \chi_{\Sch}&=1.
 \end{aligned}
 \label{eq:schwarzschild_optical_benchmark}
\end{equation}
For the half-orbit sequences in Eq.~\eqref{eq:half_orbit_compression}, these reference values give $\gamma_{\mathrm{half}}^{\Sch}=\pi$ and $w_{n+1}/w_n\to e^{-\pi}\simeq0.0432$.

\subsubsection{Mass dependence and radial-branch comparisons at fixed \texorpdfstring{$\kappa$}{kappa}}
\label{subsec:chi_fixed_kappa}
To determine whether the monotonic growth of the shadow radius is accompanied by a monotonic change in photon-orbit instability, we now examine the right panels of Fig.~\ref{fig:fixed_kappa}.
The comparison follows $\chi$ along the same branches at fixed $\kappa=100$ and $1000$ and fixed $\lambda$, with $\chi=1$ for Schwarzschild black holes as the reference.

Both radially stable branches show a nonmonotonic dependence on mass, with the extrema quoted below taken from the sampled sequences.
For $\kappa=100$, $\chi$ starts at approximately $1.402$, reaches a maximum of $1.411$ near $M/\lambda\simeq0.05939$, and then falls below unity near $M/\lambda\simeq0.085$.
The index then decreases to a minimum of approximately $0.741$ near $M/\lambda\simeq0.09417$, before recovering to $0.981$ at the high-mass end.
For $\kappa=1000$, $\chi$ starts at approximately $1.027$, rises to a maximum of $1.104$ near $M/\lambda\simeq0.02828$, and falls below unity near $M/\lambda\simeq0.042$.
The minimum is approximately $0.9396$ near $M/\lambda\simeq0.05060$, followed by an increase to $0.9894$ at the high-mass end.

The shadow radius and photon-orbit index therefore capture different aspects of the scalarized geometry.
By Eq.~\eqref{eq:optical_height_and_curvature}, the monotonic growth of $\bcrit/M$ corresponds to a decreasing dimensionless barrier height $M^2U_{\mathrm{opt}}(\rph)$.
At the minimum of $\chi$ for $\kappa=100$, the shadow is about $3.55\%$ smaller than the shadow of a Schwarzschild black hole, whereas $\chi$ is about $25.9\%$ below unity.
At the corresponding minimum for $\kappa=1000$, the departures are approximately $1.33\%$ and $6.04\%$, respectively.
These examples show that a modest size difference can accompany a more pronounced change in near-critical photon dynamics.

Where $\chi>1$, neighboring rays separate more rapidly per unit orbital angle than around Schwarzschild black holes.
Where $\chi<1$, neighboring rays separate more slowly and the leading logarithmic increase in the accumulated orbital angle is stronger.
The positive minima imply that the circular photon orbits remain unstable throughout both radially stable segments.
For the half-orbit image sequence in Eq.~\eqref{eq:half_orbit_compression}, the decrease of $\chi$ also implies weaker compression between successive high-order bands.

The right panels of Fig.~\ref{fig:unstable_optics} provide the corresponding mass comparison for the three radially unstable segments identified in Section~\ref{subsec:fixed_kappa}.
In contrast to the larger excursions on the radially stable branches, all three unstable segments have photon-orbit indices close to unity.
The low-charge $\kappa=100$ branch has $0.9987\lesssim\chi\lesssim1$, the short high-mass continuation at the same coupling has $\chi\simeq0.984\text{--}0.988$, and the lower-charge $\kappa=1000$ branch has $0.9910\lesssim\chi\lesssim1$.

At the representative matched masses $M/\lambda=0.05365$ and $0.05550$, the radially stable solutions have larger photon-orbit indices than their unstable counterparts at $\kappa=100$, but smaller indices at $\kappa=1000$.
The ordering of shadow sizes therefore does not fix the ordering of $\chi$.
The unstable solutions admit growing radial perturbations despite their optical indicators being close to those of Schwarzschild black holes: $\chi$ describes null-geodesic separation on a given metric, whereas background stability concerns perturbations of the metric and scalar field.
Neither a shadow resembling that of a Schwarzschild black hole nor $\chi\simeq1$ establishes radial stability.
Over the sampled unstable segments, the largest departure of $\chi$ from unity is only about $1.6\%$.
Together with the small shadow-size differences in Section~\ref{subsec:fixed_kappa}, this confirms that the unstable branches provide much weaker departures from the Schwarzschild optical indicators than the pronounced variations on the stable branches.

The comparison above describes the behavior along each branch over its own mass interval.
Because these intervals differ, the smaller excursion of $\chi$ at $\kappa=1000$ does not by itself establish how changing $\kappa$ affects a black hole of a specified mass.
We address that question using matched ADM masses in Section~\ref{subsec:chi_matched_mass}.

\subsubsection{Coupling dependence at matched ADM mass}
\label{subsec:chi_matched_mass}
For the coupling dependence of near-critical photon dynamics, we return to the right panels of Fig.~\ref{fig:matched_mass}.
The comparison uses radially stable backgrounds at $\kappa=100,200,\ldots,1000$ and the same sixteen matched masses in $0.05365\leq M/\lambda\leq0.06450$ as the shadow-size comparison in Section~\ref{subsec:matched_mass}, with $\lambda$ fixed.
The upper panel follows $\chi$ with mass at each coupling, and the lower panel isolates its coupling dependence at $M/\lambda=0.05365$, $0.06000$, and $0.06450$.

At each matched mass, we find that $\chi$ first decreases and then increases as $\kappa$ grows, with its smallest sampled value occurring at an intermediate coupling.
For $M/\lambda=0.05365$, $0.06000$, and $0.06450$, these minima occur at $\kappa=700$, $500$, and $400$, with $\chi\simeq0.9286$, $0.9108$, and $0.8980$, respectively.
Thus the coupling associated with the smallest index shifts toward lower values as the mass increases.
%These locations refer to the discrete coupling grid rather than precisely determined extrema of a continuous family.
%A comparison of $\kappa=100$ and $1000$ alone would miss this intermediate suppression and subsequent recovery of $\chi$.
The mass sequences in the upper-right panel also cross the reference value for Schwarzschild black holes for some couplings.
For $\kappa=100$ and $200$, all sampled values satisfy $\chi>1$, whereas for $\kappa=500,600,\ldots,1000$ they satisfy $\chi<1$.
For $\kappa=300$ and $400$, the crossings of $\chi=1$ are bracketed by $0.059<M/\lambda<0.060$ and $0.0540<M/\lambda<0.0543$, respectively.
%The value $\chi=1$ is the benchmark for Schwarzschild black holes, not a boundary between stable and unstable photon orbits: all indices remain positive, so the relevant circular null orbits remain unstable.

The different coupling dependences of the shadow radius and photon-orbit index distinguish two geometric effects.
By Eq.~\eqref{eq:optical_height_and_curvature}, the monotonic increase of $\bcrit/M$ corresponds to a decrease of the dimensionless capture-barrier height.
The nonmonotonic index instead reflects the combined dependence on the barrier curvature and the local metric at the photon sphere.
Near the sampled minima of $\chi$, neighboring rays separate more slowly per unit orbital angle, the logarithmic coefficient $1/\chi$ of the accumulated orbital angle is larger, and the asymptotic half-orbit width ratio $e^{-\pi\chi}$ is larger.
Higher-order image bands therefore undergo weaker compression toward the critical curve, under the fixed-source assumptions of Eq.~\eqref{eq:half_orbit_compression}.
%This geometric prediction does not determine their observed fluxes.

%We now examine how the parameters in Section~\ref{sec:geometry_results} are expressed in complete photon trajectories and geometric images.
%We retain the background selection and comparison strategy of Section~\ref{sec:geometry_results}.
%At fixed $\kappa=100$ and $1000$, the rays are computed on all backgrounds used in Section~\ref{subsec:fixed_kappa}, including the radially stable and radially unstable segments, with distinct segments kept separate.
%The matched-mass comparison uses the same radially stable backgrounds as Section~\ref{subsec:matched_mass}: $\kappa=100,200,\ldots,1000$ at sixteen masses spanning $0.05365\leq M/\lambda\leq0.06450$, with $\lambda$ fixed.
%Schwarzschild black holes provide the reference.
%The figures show selected mass slices and representative trajectories from this full sample to highlight the physical trends.

\subsection{Photon trajectories and photon-ring width}
\label{sec:rays}
\label{subsec:number_of_orbits}
\label{subsec:ray_paths}
Consider a light ray traced backward from an observer at infinity in the asymptotically flat spacetime.
We use the accumulated orbital angle $\psi$ introduced in Section~\ref{subsec:null_geodesics}, setting $\psi=0$ at the observer and increasing it along the backward trajectory.
The angular change along a radial segment follows from the conserved quantities in Section~\ref{subsec:null_geodesics}:
\begin{equation}
 \left|\frac{d\psi}{dr}\right|
 =\frac{b\sqrt{A(r)/B(r)}}{r^2\sqrt{1-b^2A(r)/r^2}}.
 \label{eq:ray_angular_integrand}
\end{equation}
For a captured ray, the total accumulated angle covers the path from infinity to the horizon.
For a scattered ray, it includes the ingoing and outgoing segments between infinity and the outer turning point.
Therefore, the total accumulated angle $\Delta\psi$ is
\begin{equation}
 \Delta\psi(b)=
 \begin{cases}
  \displaystyle\int_{r_H}^{\infty}\left|\frac{d\psi}{dr}\right|dr,
       & 0\leq b<\bcrit,\\[1.2ex]
  \displaystyle 2\int_{r_t(b)}^{\infty}\left|\frac{d\psi}{dr}\right|dr,
       & b>\bcrit,
 \end{cases}
 \label{eq:total_orbital_angle}
\end{equation}
where $r_t(b)$ is the outer turning point.
Dividing the total angle by one full revolution defines~\citep{GrallaHolzWald2019} the fractional number of orbits,
\begin{equation}
 N(b)=\frac{\Delta\psi(b)}{2\pi}.
 \label{eq:number_of_orbits}
\end{equation}

For an equatorial thin disk viewed along its normal, successive crossings of the disk plane occur at $\psi_m=(m-\tfrac12)\pi$, where $m=1,2,\ldots$ counts the crossings from the observer.
Following the geometric classification of Ref.~\citep{GrallaHolzWald2019}, we call rays with $N<3/4$ direct, those with $3/4<N<5/4$ lensed, and those with $N>5/4$ members of the photon-ring class.
These classes allow for rays to have at most one, exactly two, and at least three crossings of the plane, respectively.
%They describe the available light paths, not their brightness.
%A crossing contributes radiation only when it lies within the emitting portion of the disk.
We use $n=m-1$ for the discrete disk-image order, so $n=0$ denotes the direct image.
%One ray can carry contributions from several image orders.
%The thresholds in $N$ follow from the viewing geometry, whereas their locations in $b/M$ depend on the spacetime metric.

For the finite-radius calculations, the observer and the other endpoint of each scattered ray are placed at the common radius $r_o=200M$.
The subscript $o$ labels the resulting finite-radius quantities, $\Delta\psi_o$ and $N_o=\Delta\psi_o/(2\pi)$, with $N_o\to N$ as $r_o\to\infty$.
%The quoted classification boundaries use this finite-radius prescription.
%The captured path is terminated at $r_H(1+10^{-5})$ in the finite-domain calculation.

\begin{figure}[htbp]
 \centering
 \includegraphics[width=\linewidth]{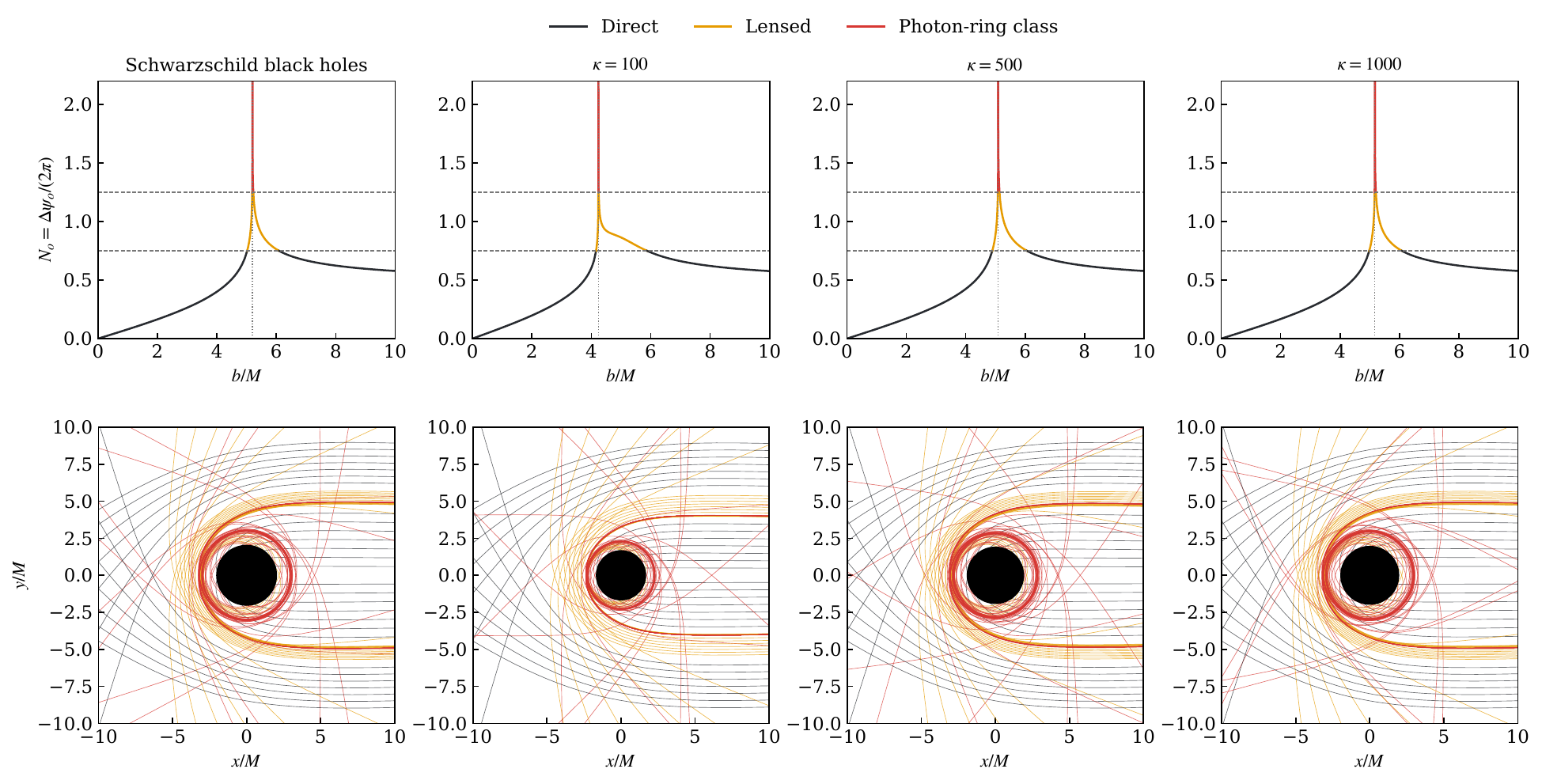}
 \caption{The upper row shows the number of orbits as a function of impact parameter, and the lower row shows selected null trajectories at $M/\lambda=0.06000$ for Schwarzschild black holes and radially stable scalarized black holes with $\kappa=100$, $500$, and $1000$.
  Rays are traced from $r_o=200M$.
 Black, gold, and red curves denote the direct, lensed, and photon-ring classes, respectively.
  In the upper panels, horizontal dashed lines mark $N_o=3/4$ and $5/4$, and vertical dotted lines mark $\bcrit/M$.
 In the lower panels, the black disk is the horizon.% and the dashed circle is the photon sphere.
 %The sampling density of trajectories does not represent intensity.
 }
 \label{fig:rays_chi}
\end{figure}

To connect the near-critical orbit analysis with complete photon paths, we compare the number of orbits and selected trajectories in Fig.~\ref{fig:rays_chi} at $M/\lambda=0.06000$ and fixed $\lambda$.
The columns show Schwarzschild black holes and radially stable scalarized backgrounds at $\kappa=100$, $500$, and $1000$.
The intermediate coupling is included because it gives the smallest sampled $\chi$ at this mass in Section~\ref{subsec:chi_matched_mass}.
The upper row plots $N_o$ against $b/M$ and identifies the direct, lensed, and photon-ring classes, while the lower row shows representative paths from those classes in Cartesian coordinates $(x,y)$ in the orbital plane.

In the upper row, $N_o$ increases toward the capture threshold from $b<\bcrit$ and decreases as $b>\bcrit$ moves away from the threshold.
The apparent spike marks $b=\bcrit$, where the ray approaches the unstable circular photon orbit and accumulates an unbounded angle.
The captured and scattered curves are separate branches, with no finite total angle at their common limiting value.
The lower row shows the corresponding behavior in real space: rays with the largest numbers of orbits remain close to $r=\rph$ before falling inward or escaping.
%The horizon radius, photon-sphere radius, and screen radius $\bcrit$ have different geometric meanings.

For $\kappa=100$, we also find a shoulder, or a locally flatter portion of the scattered-side curve, near $N_o=1$.
This feature changes the shape of $N_o(b/M)$, so the curve for this scalarized background cannot be obtained from the curve for Schwarzschild black holes by a simple shift or uniform rescaling of the impact parameter.

Table~\ref{tab:ring_boundaries} gives the four classification boundaries for the backgrounds at $M/\lambda=0.06000$ shown in Fig.~\ref{fig:rays_chi}.
The boundaries are the impact parameters at which $N_o=3/4$ or $5/4$, evaluated separately on either side of $\bcrit$.
Writing $b_{3/4}^{-}$ and $b_{5/4}^{-}$ for the captured-side thresholds and $b_{5/4}^{+}$ and $b_{3/4}^{+}$ for the scattered-side thresholds, the photon-ring class occupies
$(b_{5/4}^{-},\bcrit)\cup(\bcrit,b_{5/4}^{+})$.
The total impact-parameter width of this class is approximately $0.008683M$ for $\kappa=100$, compared with $0.038276M$ for Schwarzschild black holes and $0.042631M$ for $\kappa=1000$.
Thus the $\kappa=1000$ critical radius is close to that of Schwarzschild black holes while its photon-ring class is somewhat wider.
By comparison, the $\kappa=100$ class is substantially narrower and displaced inward.
%These are widths of a class defined by the number of orbits, not the widths of individual emitting image bands.
%The impact-parameter range of the photon ring, defined by $N>5/4$~\citep{GrallaHolzWald2019}, has been used to compare the optical properties of scalarized and Schwarzschild black holes.
To quantify this range along our mass and coupling sequences, we use~\citep{WangEtAl2024} the photon-ring width in impact-parameter space,
\begin{equation}
 W_{\rm ph}\equiv b_{5/4}^{+}-b_{5/4}^{-},
 \label{eq:photon_class_width}
\end{equation}
which measures the full interval with $N>5/4$ ($N_o>5/4$ for the finite-endpoint results).
The critical value itself is excluded from the class, but removing this single divergent ray does not change its total width.

\begin{table}[htbp]
 \centering\small
 \caption{The table lists the geometric classification boundaries at $M/\lambda=0.06000$ for endpoints at $r_o=200M$.
 Superscripts $-$ and $+$ identify the captured and scattered sides, respectively.}
 \label{tab:ring_boundaries}
 \begin{tabular}{@{}lrrrrr@{}}
 \toprule
 Background & $b_{3/4}^{-}/M$ & $b_{5/4}^{-}/M$ & $\bcrit/M$ & $b_{5/4}^{+}/M$ & $b_{3/4}^{+}/M$\\
 \midrule
 Schwarzschild black holes & 5.019378 & 5.188021 & 5.196152 & 5.226298 & 6.094220\\
 $\kappa=100$ & 4.161053 & 4.237084 & 4.238120 & 4.245767 & 5.878705\\
 $\kappa=500$ & 4.895264 & 5.082902 & 5.094518 & 5.138855 & 6.063948\\
 $\kappa=1000$ & 4.982339 & 5.162627 & 5.172237 & 5.205258 & 6.083820\\
 \bottomrule
 \end{tabular}
\end{table}

%The low-order classification intervals depend on propagation through the whole exterior geometry, whereas $\chi$ controls the limiting behavior sufficiently close to the critical orbit.

\begin{figure}[htbp]
 \centering
 \includegraphics[width=\linewidth]{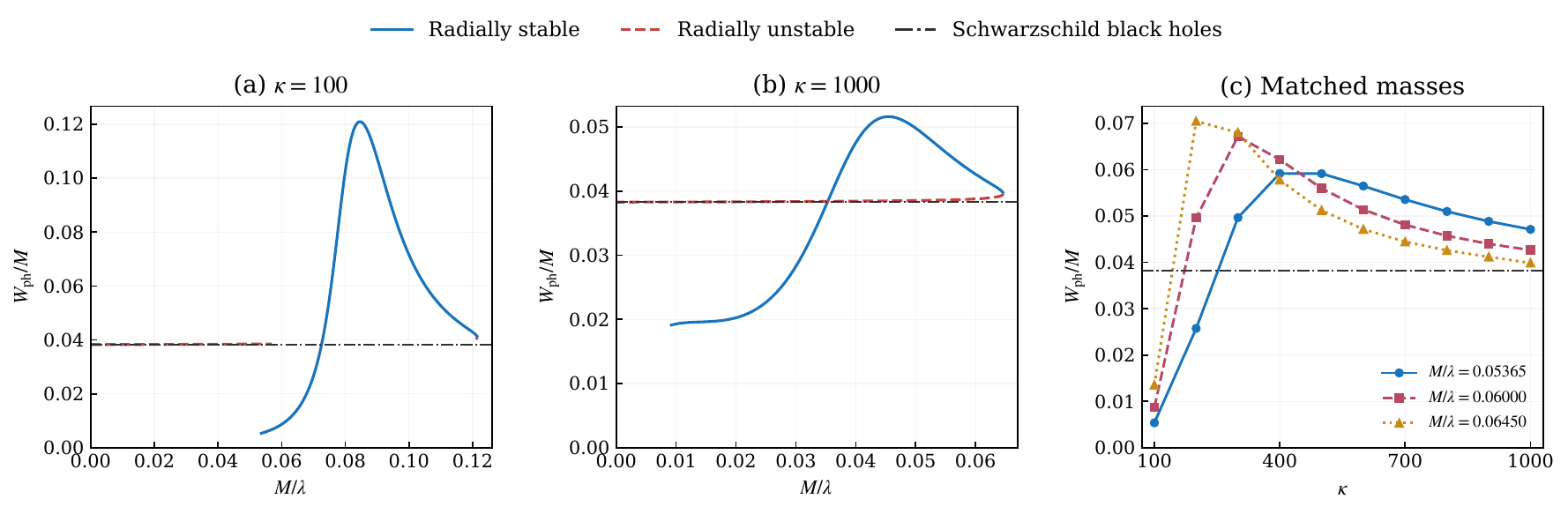}
 \caption{The photon-ring width $W_{\rm ph}/M$ is shown as a function of mass and coupling at fixed $\lambda$.
 Panels (a) and (b) vary $M/\lambda$ at $\kappa=100$ and $1000$, respectively.
 Blue solid and red dashed curves denote radially stable and radially unstable backgrounds.
 Panel (c) varies $\kappa=100,200,\ldots,1000$ for radially stable backgrounds at $M/\lambda=0.05365$, $0.06000$, and $0.06450$.
 Black dash-dotted lines give the corresponding width for Schwarzschild black holes.}
 \label{fig:ray_grid_comparison}
\end{figure}

To determine how the impact-parameter range occupied by photon-ring rays changes across the solution families, we present $W_{\rm ph}/M$ in Fig.~\ref{fig:ray_grid_comparison} at fixed $\lambda$.
Panels (a) and (b) follow the mass dependence along the radially stable and radially unstable branches at fixed $\kappa=100$ and $1000$, respectively.
%This uses the same background sequences as the shadow-size and photon-orbit-index comparison in Fig.~\ref{fig:fixed_kappa} of Section~\ref{subsec:fixed_kappa}.
Panel (c) compares radially stable solutions at $\kappa=100,200,\ldots,1000$ for each of the three fixed masses $M/\lambda=0.05365$, $0.06000$, and $0.06450$.
%These are the same matched-mass comparisons shown in the lower panels of Fig.~\ref{fig:matched_mass} in Section~\ref{subsec:matched_mass}; Schwarzschild black holes provide the reference in both comparisons.
Schwarzschild black holes provide the reference width in all three panels.

Along the radially stable $\kappa=100$ segment in panel (a), $W_{\rm ph}/M$ grows from approximately $0.00538$ at the low-mass end to a sampled maximum of $0.12094$ near $M/\lambda=0.08494$, before decreasing again.
At $\kappa=1000$, the largest sampled photon-ring width, $W_{\rm ph}/M\simeq0.05160$, occurs near $M/\lambda=0.04530$.
The radially unstable segments remain closer to the reference width $W_{\rm ph}/M\simeq0.03828$ of Schwarzschild black holes, reaching maxima of approximately $0.04091$ on the short high-mass continuation at $\kappa=100$ and $0.03957$ at $\kappa=1000$.

At the three matched masses shown in panel (c), $W_{\rm ph}/M$ varies nonmonotonically with $\kappa$, in contrast to the monotonic shadow size in Section~\ref{subsec:matched_mass}.
For example, at $M/\lambda=0.06000$, the largest sampled photon-ring width is $W_{\rm ph}/M\simeq0.06715$ at $\kappa=300$, whereas the smallest sampled $\chi$ occurs at $\kappa=500$, as shown in Fig.~\ref{fig:matched_mass}.
The largest photon-ring width therefore need not occur at the coupling with the smallest photon-orbit index.
This difference arises because $W_{\rm ph}$ depends on the accumulated angle along the full ray path, whereas $\chi$ measures the local instability near the circular photon orbit.

\subsection{Complementarity of shadow size, orbit instability, and ring width}
\label{subsec:complementarity}
The results above motivate describing the photon geometry by the pair $(\bcrit/M,\chi)$, or equivalently $(\deltash,\chi)$.
The quantities $\bcrit/M$ and $\deltash$ encode the same size information, whereas $\chi$ supplies an additional measure of near-critical photon dynamics.
By Eq.~\eqref{eq:optical_height_and_curvature}, the size fixes the dimensionless height of the capture barrier, while the index also probes the radial metric at the photon sphere.
This distinction is manifest in Fig.~\ref{fig:fixed_kappa}: the normalized shadow radius increases monotonically along each radially stable segment, but $\chi$ varies nonmonotonically.

A particularly clear illustration is provided by two radially stable backgrounds in the full-branch sample.
At $\kappa=100$ and $M/\lambda\simeq0.060208$, we obtain $\bcrit/M\simeq4.2435$ and $\chi\simeq1.4110$.
At $\kappa=1000$ and $M/\lambda\simeq0.009973$, the corresponding values are $\bcrit/M\simeq4.2398$ and $\chi\simeq1.0292$.
Their normalized shadow radii $\bcrit/M$ differ by less than $0.1\%$, yet their photon-orbit indices differ substantially.
The comparison shows that nearly equal normalized shadow radii can coexist with distinctly different photon-orbit instability.
The difference in $\chi$ has a direct consequence for the high-order image hierarchy.
For a fixed emitting radial interval and successive image orders separated by a half orbit, Eq.~\eqref{eq:half_orbit_compression} predicts width ratios approaching $0.01188$ and $0.03943$ for these two backgrounds, respectively.
The width ratio for the $\kappa=1000$ background is about $3.32$ times that for the $\kappa=100$ background, corresponding to weaker compression toward the critical curve.
Similarly, the leading logarithmic coefficient $1/\chi$ in Eq.~\eqref{eq:near_critical_angle} is larger for the $\kappa=1000$ background.
Near the critical curve, reducing the fractional impact-parameter offset by a fixed factor consequently produces a larger increase in the accumulated orbital angle, even though the two critical radii in units of mass are almost identical.

The same complementarity is present in the controlled branch comparison at fixed mass.
For $\kappa=1000$ and $M/\lambda=0.05365$, the radially stable solution has a shadow radius only $0.921\%$ smaller than that of the unstable solution, but its index is $5.36\%$ smaller.
Their asymptotic half-orbit width ratios are approximately $0.05142$ and $0.04347$, respectively, so the stable background has a width ratio about $18.3\%$ larger.
%These are geometric predictions for high-order bands, not ratios of observed fluxes or statements about the widths of the first few image orders.

Together, shadow size and $\chi$ specify the position of the critical curve and the asymptotic manner in which image bands approach it.
The photon-ring width $W_{\rm ph}/M$ adds a finite-angle measure: it specifies the impact-parameter interval occupied by rays in the photon-ring class defined in Section~\ref{subsec:ray_paths}.
These complementary quantities provide the geometric basis for the radiative images in Section~\ref{sec:imaging}, where the emissivity, emitter motion, and emitting region determine how those light paths contribute to the observed intensity.

\section{Spherical Emission and Thin-Disk Images}\label{sec:imaging}
We focus on the radially stable scalarized backgrounds to examine how the geometric departures found in Section~\ref{sec:geometry_results} appear under different emission models.
The radially unstable branches serve as geometric controls in Figs.~\ref{fig:fixed_kappa} and~\ref{fig:unstable_optics}, where their shadow radii and photon-orbit indices remain close to those of Schwarzschild black holes.
We describe the images using Cartesian impact-parameter coordinates $(b_x,b_y)$, with $b=\sqrt{b_x^2+b_y^2}$ measuring distance from the image center.
%We now combine the photon propagation studied in Section~\ref{sec:geometry_results} with prescribed emission models to determine the observed brightness distributions.

%\subsection{Emission models and radiative transfer}
%\label{subsec:transfer}
%For the radiative calculations in this section, we compare radially stable backgrounds at $\kappa=100$ and $1000$ and $M/\lambda=0.05365$, $0.05550$, $0.06000$, and $0.06450$, together with the same-mass Schwarzschild black holes, keeping $\lambda$ fixed.
%These four masses and two couplings form a representative subset of the matched-mass comparison in Fig.~\ref{fig:matched_mass}, Section~\ref{subsec:matched_mass}.
%For the ISCO-truncated circular disk, we additionally vary mass along the full radially stable sequences at fixed $\kappa=100$ and $1000$, as in Fig.~\ref{fig:fixed_kappa}, and compare $\kappa=100,200,\ldots,1000$ at sixteen fixed masses spanning $0.05365\leq M/\lambda\leq0.06450$, as in Fig.~\ref{fig:matched_mass}.
%We consider two idealized emission geometries on the same backgrounds: an optically thin spherical volume and a transparent, geometrically thin equatorial disk viewed face-on.

\subsection{Spherical emission models}\label{subsec:spherical_emission}
We model the emitting matter as an optically thin, spherically symmetric distribution with negligible backreaction on the black-hole spacetime.
Here, ``optically thin'' means that photons traverse the emitting region with negligible absorption or scattering.
Emission is isotropic in the local rest frame, and we consider both static emitters and matter in radial free fall from rest at infinity.

For a photon with four-momentum $k^\mu$, the frequency shift between emission and observation is characterized by the dimensionless factor
\begin{equation}
 \mathcal{Z}\equiv\frac{\nu_o}{\nu_e}
 =\frac{-k_\mu u_o^\mu}{-k_\mu u_e^\mu},
 \qquad u_o^\mu=A_o^{-1/2}(\partial_t)^\mu,
 \label{eq:radiative_frequency_shift}
\end{equation}
where $\nu_o$ and $\nu_e$ denote the photon frequencies measured in the local rest frames of the observer ($o$) and emitter ($e$), while $u_o^\mu$ and $u_e^\mu$ are their respective four-velocities.
The observer is static at radius $r_o$, where the metric function takes the value $A_o=A(r_o)$.
We consider bolometric intensities, meaning intensities integrated over all photon frequencies.
Neglecting absorption and scattering, the observed intensity is therefore
\begin{equation}
 I_o=\int_\gamma \mathcal{Z}^4 j_{\mathrm{e}}\,d\ell_e,
 \label{eq:bolometric_transfer}
\end{equation}
where $I_o$ is the radiative energy received per unit time, projected area, and solid angle, integrated over all observed frequencies.
The comoving emissivity $j_{\mathrm{e}}$ is the energy emitted per unit proper time, proper volume, and solid angle, integrated over all frequencies in the emitter's rest frame.
The integration path $\gamma$ is the photon's null geodesic reaching the observer, and $d\ell_e$ is the spatial path-length element measured in the emitter's local rest frame.
%The fourth power of $\mathcal{Z}$ combines the cubic transformation of intensity per unit frequency with the additional factor from frequency integration, $d\nu_o=\mathcal{Z}\,d\nu_e$.
%Both $\mathcal{Z}$ and the emitter-frame path length change when the emitting material moves.

%We use $y=r/M$ and $\beta=b/M$, with each background's actual ADM mass.
%The common outer emission boundary is $30M$, and the reference emissivity is normalized at $6M$.
%The baseline radial power is $p=2$, with $p=1$ and $3$ used to vary the radial weighting.
%Images use impact-parameter coordinates $(b_x/M,b_y/M)$ and the observer at $200M$.
%We compare intensities within each emission family using a common local normalization; this does not fix a common mass accretion rate or total luminosity.
%The baseline comparison at four matched masses contains 12 spacetime cases and 18 emission prescriptions per case.
%The extended disk comparison keeps $p=2$ and contrasts the ISCO inner edge with the common $6M$ inner edge on each background.

For the spherical volume, the isotropic comoving emissivity is
\begin{equation}
 j_{\mathrm{e}}(r)=j_0\left(\frac{r}{r_{\mathrm{ref}}}\right)^{-p}
 \mathbf{1}_{[r_{\mathrm{in}},r_{\mathrm{out}}]}(r),
 \label{eq:spherical_emissivity}
\end{equation}
where $r_{\mathrm{in}}$ and $r_{\mathrm{out}}$ are the inner and outer radii of the emitting spherical shell.
The independent reference radius $r_{\mathrm{ref}}$ sets the normalization, with $j_0$ the local emissivity there and $p$ the radial power-law index.
The spherical-emission results below use $r_{\mathrm{ref}}=6M$ and the same $j_0$ across the compared backgrounds and emitter motions.
Thus the normalization radius is distinct from the inner boundary of the emitting region.
The indicator function $\mathbf{1}_{[r_{\mathrm{in}},r_{\mathrm{out}}]}(r)$ equals $1$ for $r_{\mathrm{in}}\leq r\leq r_{\mathrm{out}}$ and $0$ otherwise.
%The baseline has $y_{\mathrm{in}}=(r_H/M)(1+10^{-3})$ and $y_{\mathrm{out}}=30$.
%We also consider a common inner boundary at $6M$ and an outer boundary at $60M$ as separate variations.
%Intensities are reported in units of $j_0M$.
%The volume integral has no discrete disk-crossing decomposition.
To express the transfer factors compactly, we define $\mathcal P(r,b)=1-b^2A(r)/r^2$, the dimensionless radial factor in Eq.~\eqref{eq:null_radial_motion}.
Allowed ray segments have $\mathcal P\geq0$, and a radial turning point satisfies $\mathcal P=0$.
For static emitters, the frequency shift and local path length are
\begin{equation}
 \mathcal{Z}_{\mathrm{stat}}=\sqrt{\frac{A(r)}{A_o}},
 \qquad
 d\ell_{\mathrm{stat}}
 =\frac{|dr|}{\sqrt{B(r)\mathcal P(r,b)}}.
 \label{eq:static_spherical_transfer}
\end{equation}
Scattered rays receive emission along both radial segments, but captured backward rays receive emission along the single segment outside the horizon.
Thus the region $b<\bcrit$ can have nonzero intensity even though rays entering it from the observer do not return to the outer boundary.

For radial free fall from rest at infinity, the emitter has conserved energy per unit rest mass $-u_{e,t}=1$ and four-velocity
\begin{equation}
 u_e^t=\frac{1}{A(r)},\qquad
 u_e^r=-\sqrt{B(r)\left(\frac{1}{A(r)}-1\right)}.
 \label{eq:infall_four_velocity}
\end{equation}
Let $s=+1$ denote a future-directed outgoing photon segment and $s=-1$ an ingoing segment.
The dimensionless factor $\mathcal D_s(r,b)=1+s\sqrt{[1-A(r)]\mathcal P(r,b)}$ accounts for the direction dependence introduced by the infalling emitter's motion.
Using the subscript $\mathrm{ff}$ for free fall, the frequency shift and path-length transformation are
\begin{equation}
 \mathcal{Z}_s=\frac{A(r)}{\sqrt{A_o}\,\mathcal D_s(r,b)},\qquad
 \frac{d\ell_{\mathrm{ff}}}{d\ell_{\mathrm{stat}}}
 =\frac{\mathcal D_s(r,b)}{\sqrt{A(r)}}.
 \label{eq:infall_transfer_factors}
\end{equation}
Substituting these factors and Eq.~\eqref{eq:static_spherical_transfer} into Eq.~\eqref{eq:bolometric_transfer} gives
\begin{equation}
 I_o^{\mathrm{ff}}(b)
 =\frac{1}{A_o^2}\sum_{s=\pm1}\int_{\gamma_s}
 \frac{A(r)^{7/2}j_{\mathrm{e}}(r)}
 {\mathcal D_s(r,b)^3\sqrt{B(r)\mathcal P(r,b)}}\,|dr|,
 \label{eq:infall_bolometric_intensity}
\end{equation}
where $\gamma_s$ denotes the part of a ray with direction $s$ inside the emitting region.
Both signs contribute for scattered rays, whereas captured backward rays include only $s=+1$, corresponding to photons traveling outward toward the observer.
%The factor $\mathcal D_s^{-3}$ results from combining $\mathcal{Z}_s^4$ with the emitter-frame path element.
The additional redshift of radiation escaping against the ingoing motion suppresses the central brightness.

In both spherical emission models, nonzero emission near the photon sphere leads~\citep{GrallaHolzWald2019} to a logarithmic divergence of the ideal intensity at the critical curve, as near-critical rays accumulate emission over an arbitrarily long optical path.
This divergence is integrable, so averaging over a finite image area gives a finite intensity.
We therefore compare the peak intensities of static and freely infalling spherical emission after area averaging over annuli of fixed radial width $\Delta b/M=0.01$ in the image plane.
We denote the maximum of this area-averaged intensity by $I_{\mathrm{pk}}$.
%These finite peaks characterize the images at the specified averaging scale.
%Here, they do not represent a finite pointwise maximum of the ideal intensity, and their values depend on the averaging width.

\begin{figure}[htbp]
 \centering
 \includegraphics[width=\linewidth]{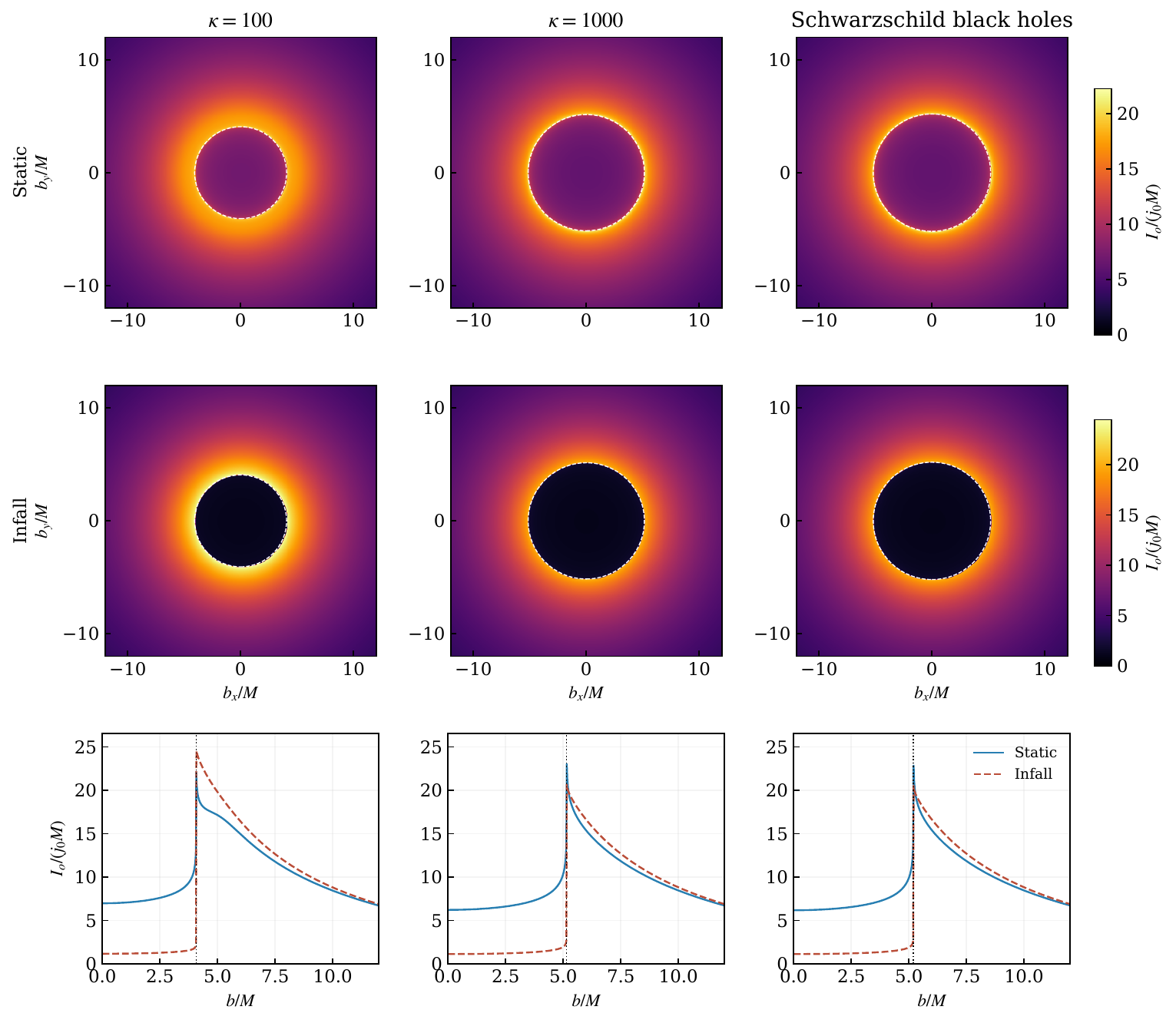}
 \caption{The images compare spherical emission at $M/\lambda=0.05365$, with $p=2$ and emission extending from just outside $r_H$ to $30M$.
 From left to right, the columns show radially stable scalarized backgrounds at $\kappa=100$ and $1000$, followed by Schwarzschild black holes, at fixed $\lambda$.
  The upper and middle rows show static emitters and radial free fall, respectively, with a common linear intensity scale across backgrounds within each row.
  White dashed circles mark $\bcrit$.
 The lower row compares the static and infalling radial profiles, where dotted vertical lines mark $\bcrit/M$.
  Intensities use the same comoving normalization $j_0$ at $r_{\mathrm{ref}}=6M$ in the two spherical emission models.}
 \label{fig:sphere_images}
\end{figure}

To examine how the geometric shadow features appear under different emitter motions, we compare static and freely infalling spherical emission in Fig.~\ref{fig:sphere_images} at fixed $\lambda$ and $M/\lambda=0.05365$.
The columns show radially stable backgrounds at $\kappa=100$ and $1000$, followed by Schwarzschild black holes, with the same $p=2$ emissivity extending from just outside $r_H$ to $30M$ and the same normalization $j_0$ at $r_{\mathrm{ref}}=6M$.
The first two rows display the static and infalling images, respectively, and the lower row compares their radial intensity profiles.

Both emission models retain the inward displacement of the bright ring near the critical curve for $\kappa=100$, whose shadow radius is $21.65\%$ smaller than that of a Schwarzschild black hole of the same mass.
For $\kappa=1000$, the shadow radius is only $0.968\%$ smaller, and the images remain close to the reference images for Schwarzschild black holes.
Changing the emitter motion on a fixed background leaves the geometric critical curve unchanged but modifies the intensity contrast across the image.

Comparing the central intensities with the area-averaged peaks, we obtain central-to-peak ratios of $0.3154$ and $0.0477$ for static emission and infall, respectively, at $\kappa=100$.
The corresponding ratios are $0.2703$ and $0.0550$ for Schwarzschild black holes, and $0.2701$ and $0.0547$ for $\kappa=1000$.
Radial infall therefore produces a darker center relative to the bright ring on each background, consistent with the additional redshift of radiation escaping against the infalling matter.

\FloatBarrier

\begin{figure}[htbp]
 \centering
 \includegraphics[width=\linewidth]{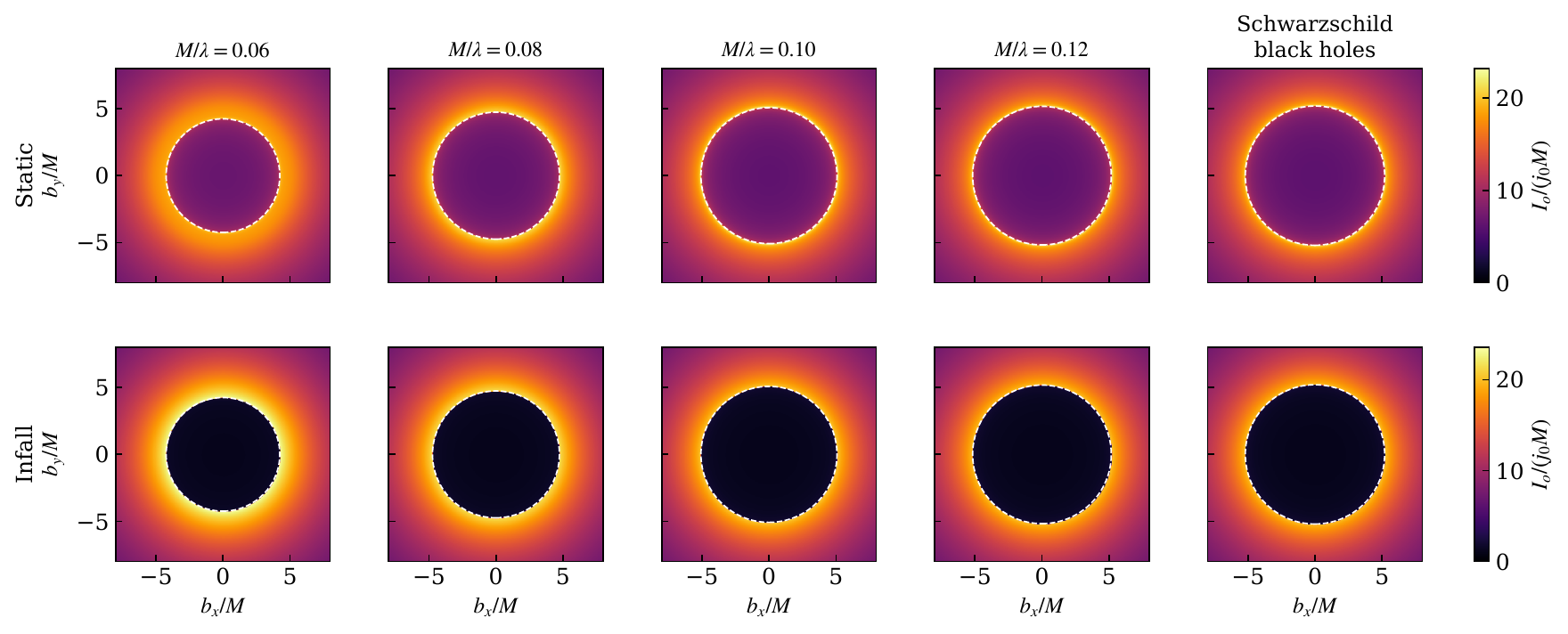}
 \caption{The spherical-emission images follow the mass dependence along the radially stable branch at fixed $\kappa=100$ and fixed $\lambda$.
 The first four columns show scalarized backgrounds at the evenly spaced masses $M/\lambda=0.06$, $0.08$, $0.10$, and $0.12$, and the last column shows Schwarzschild black holes.
 The upper and lower rows show static emitters and radial free fall from rest at infinity, respectively.
  We keep $p=2$, emission from just outside $r_H$ to $30M$, and the comoving normalization $j_0$ at $r_{\mathrm{ref}}=6M$ fixed.
 Each row shares a linear color scale for $I_o/(j_0M)$ and white dashed circles mark $\bcrit$.}
 \label{fig:sphere_mass_k100}
\end{figure}

The pronounced image differences at $\kappa=100$ motivate following how the scalarized images approach those of Schwarzschild black holes as mass increases.
For this purpose, we present the mass sequence along the radially stable $\kappa=100$ branch in Fig.~\ref{fig:sphere_mass_k100}, keeping $\lambda$ and the emission prescriptions fixed.
%This is the same fixed-coupling comparison as in Fig.~\ref{fig:fixed_kappa}, now applied to the static and freely infalling emission models with the same emissivity prescription as in Fig.~\ref{fig:sphere_images}.
The first four columns show the evenly spaced masses $M/\lambda=0.06$, $0.08$, $0.10$, and $0.12$, while the rightmost column shows Schwarzschild black holes.
The upper and lower rows compare static emission and radial free fall, respectively, using the same $p=2$ emissivity from just outside $r_H$ to $30M$ as in Fig.~\ref{fig:sphere_images}.
%With the same emission prescription expressed in units of $M$, the images of Schwarzschild black holes in coordinates $(b_x/M,b_y/M)$ and intensity units $j_0M$ are independent of mass.
%This normalization therefore allows us to identify changes in the scalarized images relative to a common reference.

Across the four scalarized backgrounds, the critical radius $\bcrit/M$ increases from $4.2381$ to $5.1695$, and the bright ring moves outward toward the reference location in both models.
The shadow radius is smaller than that of a Schwarzschild black hole with the same ADM mass by approximately $18.44\%$, $8.91\%$, $2.37\%$, and $0.51\%$, respectively.

\begin{figure}[htbp]
 \centering
 \includegraphics[width=\linewidth]{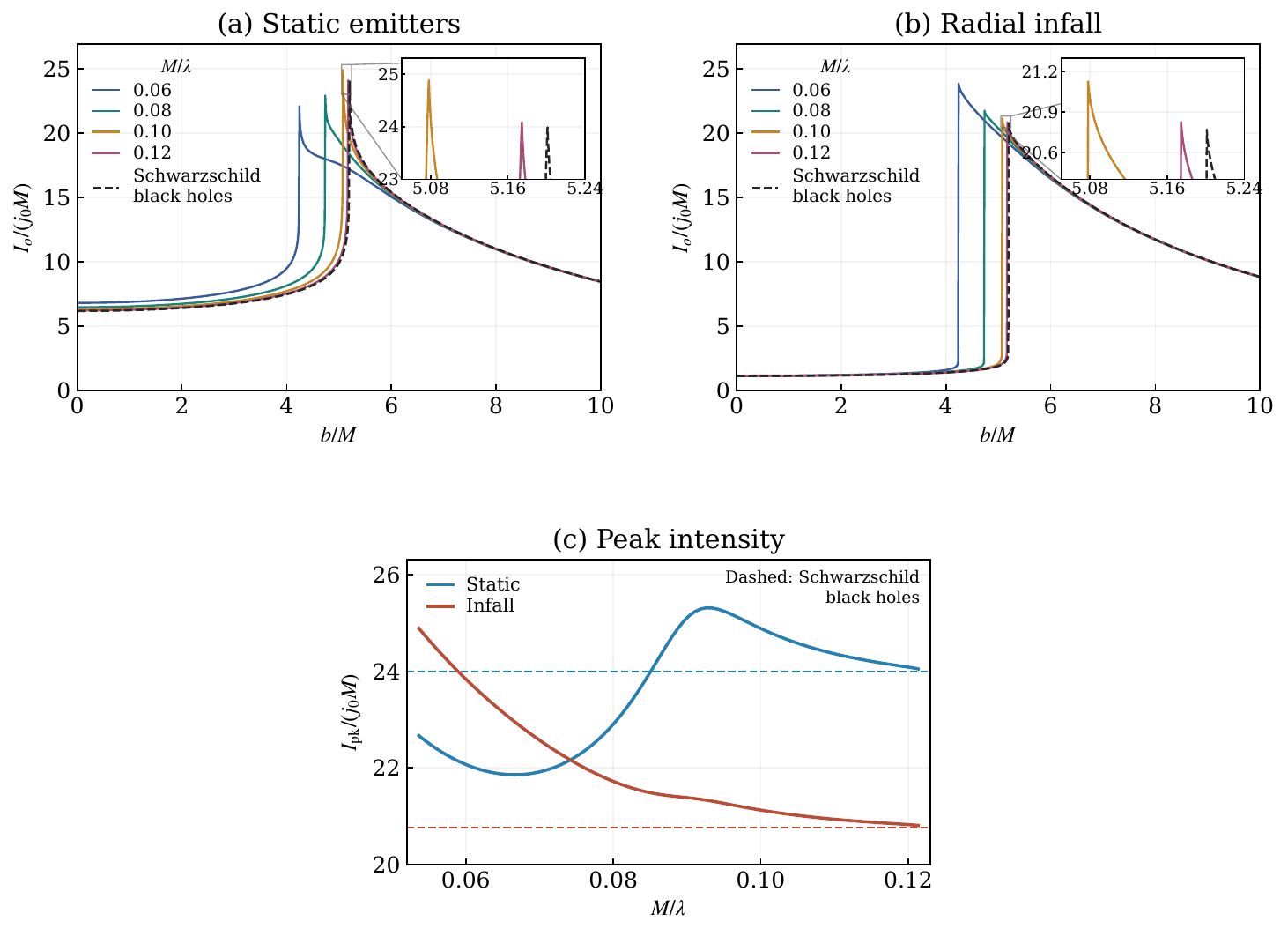}
 \caption{The radial intensity profiles and peak intensities are compared for spherical emission along the radially stable branch at fixed $\kappa=100$ and fixed $\lambda$.
 Panels (a) and (b) compare static emission and radial infall, respectively, at $M/\lambda=0.06$, $0.08$, $0.10$, and $0.12$, using the same backgrounds and emission prescriptions as Fig.~\ref{fig:sphere_mass_k100}.
 Black dashed curves show Schwarzschild black holes, and the insets enlarge the peaks for $M/\lambda=0.10$, $0.12$, and Schwarzschild black holes with the same colors and line styles.
 Panel (c) shows how the peak intensity $I_{\mathrm{pk}}$ varies with mass, with the peak defined as the maximum of the area-averaged intensity over annuli of width $0.01M$.
 Horizontal dashed lines give the corresponding mass-independent peaks for Schwarzschild black holes.}
 \label{fig:sphere_mass_profiles_k100}
\end{figure}

To quantify the brightness changes accompanying the outward motion of the ring, we present the radial intensity profiles and peak intensities in Fig.~\ref{fig:sphere_mass_profiles_k100} at fixed $\kappa=100$ and $\lambda$.
Panels (a) and (b) compare static emission and radial infall at the same four masses, $M/\lambda=0.06$, $0.08$, $0.10$, and $0.12$, as the images in Fig.~\ref{fig:sphere_mass_k100}.
Panel (c) follows the peak intensity over the radially stable mass sequence, with the peaks defined by area averaging over annuli of width $0.01M$.
The reference curves and horizontal lines show Schwarzschild black holes under the same emission and averaging prescriptions.

The profiles in Fig.~\ref{fig:sphere_mass_profiles_k100}(a,b) show that the distinction in central contrast between the two emission models persists across the four masses.
%Comparing the central intensity with the intensity peak, the ratios range from $0.2534$ to $0.3084$ for static emitters and from $0.0487$ to $0.0543$ for radial infall among the displayed scalarized backgrounds.
Among the four displayed masses, the peak intensity in units of $j_0M$ decreases as mass increases for freely infalling emitters, whereas it first increases and then decreases for static emitters.
The insets resolve the smaller differences between the peaks at $M/\lambda=0.10$, $0.12$, and the reference peaks for Schwarzschild black holes.
%Since the ideal optically thin intensity diverges at the critical curve, we compare finite peaks obtained by averaging over annuli of width $0.01M$.

Over the full mass sequence in Fig.~\ref{fig:sphere_mass_profiles_k100}(c), we find an additional low-mass minimum for static emitters: $I_{\mathrm{pk}}/(j_0M)$ falls to approximately $21.86$ near $M/\lambda\simeq0.0667$, rises to a maximum of approximately $25.31$ near $M/\lambda\simeq0.0926$, and then decreases.
This nonmonotonic trend persists for averaging widths of $0.005M$ and $0.02M$.
For radial infall, the peak decreases from approximately $24.88$ to $20.81$.
The horizontal reference lines give $I_{\mathrm{pk}}/(j_0M)\simeq23.99$ for static emission and $20.77$ for radial infall around Schwarzschild black holes.
The static-emission peak crosses its reference, whereas the infall peak approaches its reference from above.
Thus, the monotonic approach of the shadow size to the value for Schwarzschild black holes does not require a monotonic approach of the brightness peak, which also depends on the emitter motion and radiative transfer.

\FloatBarrier

\subsection{Thin-disk motion, orbital structure, and inner edges}
\label{subsec:thin_disk}

We next consider a transparent, geometrically thin equatorial disk viewed face-on, along its normal.
Here geometric thinness means that the vertical thickness is neglected and emission is assigned to intersections of the ray with the equatorial plane.
We prescribe the comoving bolometric surface intensity
\begin{equation}
 I_{\mathrm{em}}(r)=I_0\left(\frac{r}{r_{\mathrm{in}}}\right)^{-p}
 \mathbf{1}_{[r_{\mathrm{in}},30M]}(r),
 \qquad
 I_o(b)=\sum_m \mathcal{Z}_m^4 I_{\mathrm{em}}\bigl(r_m(b)\bigr).
 \label{eq:disk_intensity_sum}
\end{equation}
Here $I_0$ is the comoving bolometric surface intensity at the disk inner edge $r_{\mathrm{in}}$, $p$ determines the radial decline of $I_{\mathrm{em}}(r)$, and the transfer function $r_m(b)$ is the radius at which the backward ray first reaches the accumulated angle $\psi_m=(m-\tfrac12)\pi$ for crossing number $m$.
The factor $\mathcal Z_m$ is the frequency shift at that crossing, evaluated for the chosen emitter motion.
Only intersections inside the emitting disk contribute to the sum.
This is a specified emitting-surface model, and no additional incidence-angle factor from a finite-thickness volume model is imposed.
We consider both a static disk and a disk whose emitting matter follows circular geodesics.
The static emitter uses $\mathcal{Z}_{\mathrm{stat}}$ in Eq.~\eqref{eq:static_spherical_transfer}.
For a circular geodesic emitter, let $\Omega_{\mathrm{circ}}=d\varphi/dt$ denote the matter's orbital angular velocity. Then
\begin{equation}
 \Omega_{\mathrm{circ}}^2=\frac{A'}{2r},\qquad
 u_e^t=\frac{1}{\sqrt{A-rA'/2}},\qquad
 \mathcal{Z}_{\mathrm{circ}}=\sqrt{\frac{A-rA'/2}{A_o}}.
 \label{eq:circular_disk_redshift}
\end{equation}
The expression for $\mathcal{Z}_{\mathrm{circ}}$ uses the vanishing photon angular momentum about the disk normal, $L_z=0$, for an observer on the disk axis.

An important concept in circular geodesic motion is the innermost stable circular orbit (ISCO).
It marks the inner limit of stable circular motion along the orbit sequence approached from large radius.
For timelike circular orbits, $A'>0$ and $2A-rA'>0$ are required in the static exterior.
Radial stability can be expressed~\citep{Song2021ISCO} in terms of the stability function
\begin{equation}
 \mathcal S(r)\equiv3AA'+rAA''-2r(A')^2,
 \label{eq:disk_msco_condition}
\end{equation}
where timelike circular orbits exist, $\mathcal S>0$ indicates radial stability, $\mathcal S<0$ indicates radial instability, and $\mathcal S=0$ gives the marginally stable radii.
The function has dimensions of inverse length, and its relation to the timelike effective potential is given in Appendix~\ref{app:msco}.
When multiple marginally stable circular orbits (MSCOs) separate disconnected stable regions, this prescription selects the outer ISCO, defined as the inner boundary of the stable region continuously connected to spatial infinity, as also adopted~\citep{DeliyskiEtAl2026} in imaging studies of black holes with synchronized scalar hair.
We denote this radius by $\risco$ throughout this work, and it need not coincide~\citep{DelgadoEtAl2022} with the globally innermost stable circular orbit.
The disconnected inner stable regions are assumed not to contribute to the emission in our model.

For the static disk model, we set $r_{\mathrm{in}}=6M$ for every background.
For the rotating disk model, we consider two choices: the same inner edge $r_{\mathrm{in}}=6M$, and an inner edge $r_{\mathrm{in}}=\risco$ determined separately for each spacetime.
The first choice allows static and rotating emitters to be compared over the same radial interval, while the second incorporates the orbital structure of the background.
Both choices use the same comoving intensity $I_0$ at the disk's own inner edge.

%The stability condition Eq.~\eqref{eq:disk_msco_condition} can be expressed in terms of the radial epicyclic frequency $\Omega_r$ and the circular orbital frequency $\Omega_\varphi$, both measured with respect to the asymptotically normalized time:
%\begin{equation}
% \Omega_r^2=\frac{B\mathcal S}{2Ar},\qquad
% \Omega_\varphi^2=\frac{A'}{2r},\qquad
% \mathcal R\equiv\frac{\Omega_r^2}{\Omega_\varphi^2}
% =B\left(3+\frac{rA''}{A'}-\frac{2rA'}{A}\right).
% \label{eq:timelike_frequency_ratio}
%\end{equation}
%Positive $\mathcal R$ corresponds to radial oscillations about a circular orbit, whereas negative $\mathcal R$ indicates radial instability.
%For Schwarzschild black holes, $\mathcal R=1-6M/r$.
%Since $B>0$ in the static exterior, the MSCO locations are controlled by $A$ and its radial derivatives.

\begin{figure}[htbp]
 \centering
 \includegraphics[width=\linewidth]{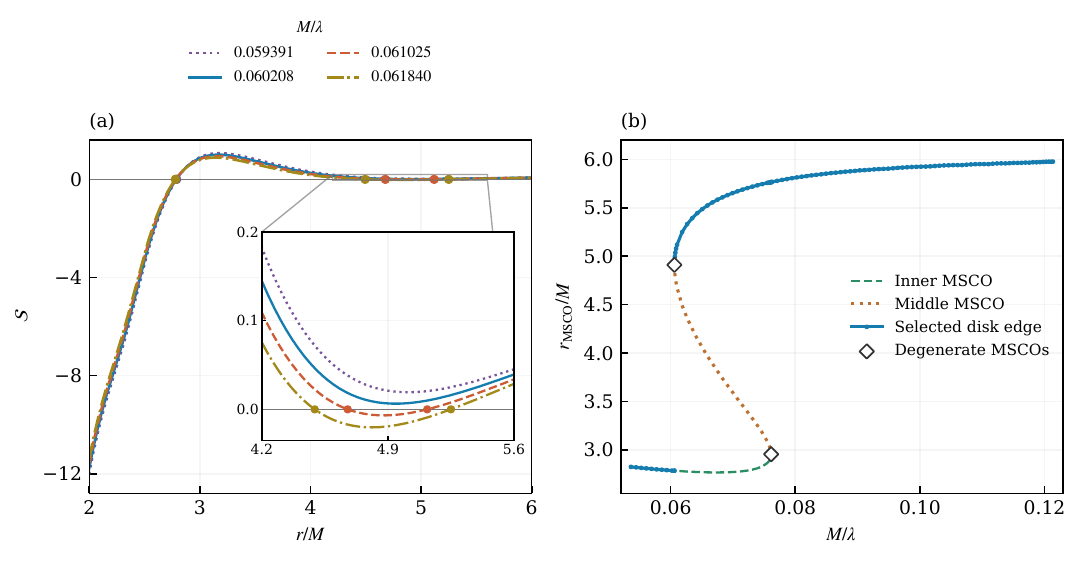}
 \caption{The figure shows circular-orbit stability and marginal radii as mass varies along the radially stable branch at fixed $\kappa=100$ and $\lambda$.
  Panel (a) shows the stability function $\mathcal S$ defined in Eq.~\eqref{eq:disk_msco_condition}, in units of $\lambda^{-1}$, over $2\leq r/M\leq6$ for the four adjacent masses listed in the legend.
  The inset enlarges $4.2\leq r/M\leq5.6$, where the outer local minimum crosses zero and produces an unstable band bounded by two additional MSCOs.
  Dots mark the MSCOs.
 Panel (b) follows all marginal radii along the radially stable background sequence: the green dashed and orange dotted curves denote the inner and middle roots, while the blue curve marks the selected disk inner edge.
  Open diamonds mark degenerate MSCOs, where two marginal radii merge and satisfy $\mathcal S=\partial_r\mathcal S=0$.}
 \label{fig:isco_metric_transition}
\end{figure}

The ISCO prescription makes the disk inner edge sensitive to changes in circular-orbit stability.
To examine the origin and mass dependence of the unusually small ISCO at $\kappa=100$, we present the stability function $\mathcal S$ and the marginally stable orbit radii in Fig.~\ref{fig:isco_metric_transition} along the radially stable branch at fixed $\lambda$.
Panel (a) compares $\mathcal S(r)$ for four neighboring masses, with an inset resolving the outer local minimum, while panel (b) follows all MSCO radii and identifies the selected disk inner edge.

The contrast with larger couplings is substantial: at $M/\lambda=0.06000$, $\risco/M\simeq2.7864$ for $\kappa=100$, compared with approximately $5.7406$, $5.9395$, and $5.9819$ for $\kappa=200$, $500$, and $1000$, respectively.
The small ISCO reflects the radial shape of the time-component metric function $A=-g_{tt}$, including its derivatives.
For example, at $r=4M$ the $\kappa=100$ value of $A$ differs from that of Schwarzschild black holes by only about $2.58\%$, while $|M^2A''|$ is smaller by about $55.7\%$.
The resulting $\mathcal S>0$ permits a stable circular orbit there, whereas $\mathcal S<0$ at the same $r/M$ for Schwarzschild black holes.

In panel (a), the outer local minimum of $\mathcal S$ crosses zero as mass increases, creating two additional MSCOs and an intervening interval with $\mathcal S<0$.
At $M/\lambda\simeq0.060208$, stable circular motion extends outward from approximately $2.79M$.
At the next background, $M/\lambda\simeq0.061025$, the three marginal radii are approximately $2.78M$, $4.68M$, and $5.12M$, and an unstable band separates the inner stable region from the stable region connected to large radius.
The selected ISCO consequently moves outward to the outermost marginal radius, because the disk model uses the stable region connected to infinity.
Panel (b) follows the resulting change from one to three MSCOs and the subsequent return to one.
The inner and middle roots merge and disappear at $M/\lambda\simeq0.0761276$ and $r/M\simeq2.95419$.
%The outer root survives this second change, so the selected disk model continues without a comparable jump.

\begin{figure}[htbp]
 \centering
 \includegraphics[width=\linewidth]{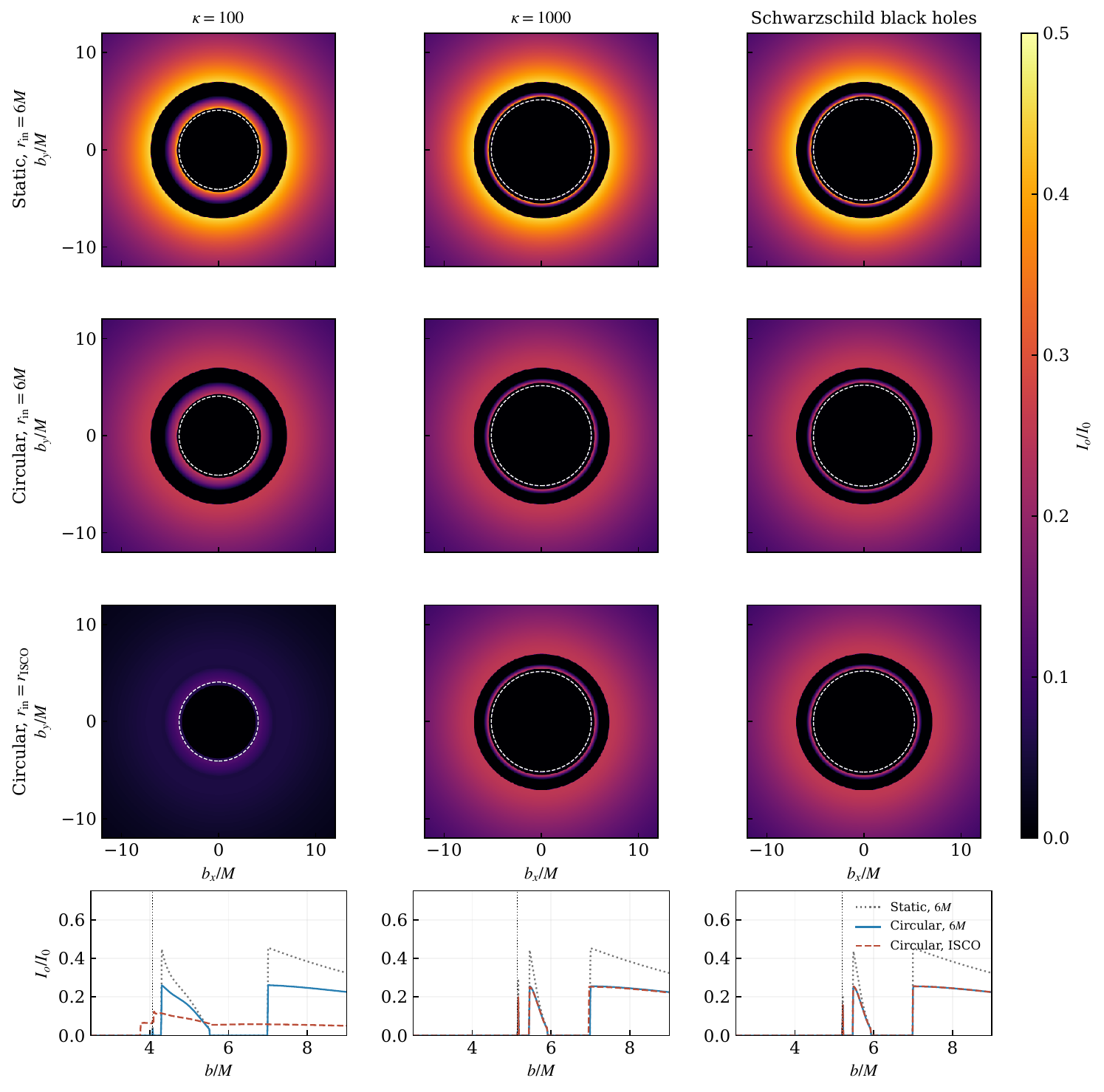}
 \caption{The images show face-on thin disks at fixed $\lambda$ and $M/\lambda=0.05365$, with $p=2$ and outer disk radius $30M$.
 From left to right, the columns show radially stable $\kappa=100$ and $1000$ backgrounds and Schwarzschild black holes.
 The first three rows show static and circular disks with $r_{\rm in}=6M$, followed by circular disks with $r_{\rm in}=\risco$.
 All images share the inner-edge intensity $I_0$ and linear color scale $0\leq I_o/I_0\leq0.5$.
 White dashed circles mark $\bcrit$.
 The fourth row shows intensity profiles as gray dotted, blue solid, and red dashed curves, respectively.
 Vertical dotted lines mark $\bcrit/M$.}
 \label{fig:disk_images}
\end{figure}

To distinguish the effects of spacetime geometry, emitter motion, and the disk inner edge, we compare the three disk prescriptions in Fig.~\ref{fig:disk_images} at fixed $\lambda$ and $M/\lambda=0.05365$.
The columns show the radially stable $\kappa=100$ and $1000$ backgrounds alongside Schwarzschild black holes.
The first two rows compare static and circular disks with the same inner edge $r_{\mathrm{in}}=6M$, while the third row places the circular disk's inner edge at the ISCO of each background.
The fourth row gives the corresponding radial intensity profiles, with the same inner-edge intensity $I_0$ and emissivity index $p=2$ used throughout the comparison.

On a given background, replacing static emitters with circularly orbiting emitters at fixed $r_{\mathrm{in}}=6M$ lowers the observed intensity without changing the photon paths or the emitting radial interval.
The dimming follows from the smaller frequency-shift factor $\mathcal{Z}_{\mathrm{circ}}$ in Eq.~\eqref{eq:circular_disk_redshift}, and is visible in both the first two image rows and the gray dotted and blue solid profiles.
Across the three backgrounds, the broad outer emission remains similar, whereas the secondary inner ring lies at smaller $b/M$ and is broader for $\kappa=100$.
The $\kappa=1000$ images and profiles are close to those for Schwarzschild black holes.

Comparing the second and third rows, we find a much stronger response to the ISCO prescription for $\kappa=100$, where the disk inner edge moves from $6M$ to $\risco\simeq2.8235M$.
The inward extension allows captured backward rays to intersect the emitting disk, producing nonzero intensity at $b<\bcrit$, as shown by the red dashed profile in the leftmost column.
The image is nevertheless dimmer overall than the circular disk with $r_{\mathrm{in}}=6M$.
Because $I_0$ is fixed at each disk's own inner edge, reducing $r_{\mathrm{in}}$ in Eq.~\eqref{eq:disk_intensity_sum} lowers the prescribed emission intensity throughout the outer region shared by the two disks.
For $\kappa=1000$, the ISCO remains close to $6M$, at $\risco/M\simeq5.9668$, so the two circular-disk images differ only slightly and their profiles nearly overlap.
For Schwarzschild black holes, $\risco=6M$, and the two circular-disk prescriptions coincide.

\begin{figure}[htbp]
 \centering
 \includegraphics[width=\linewidth]{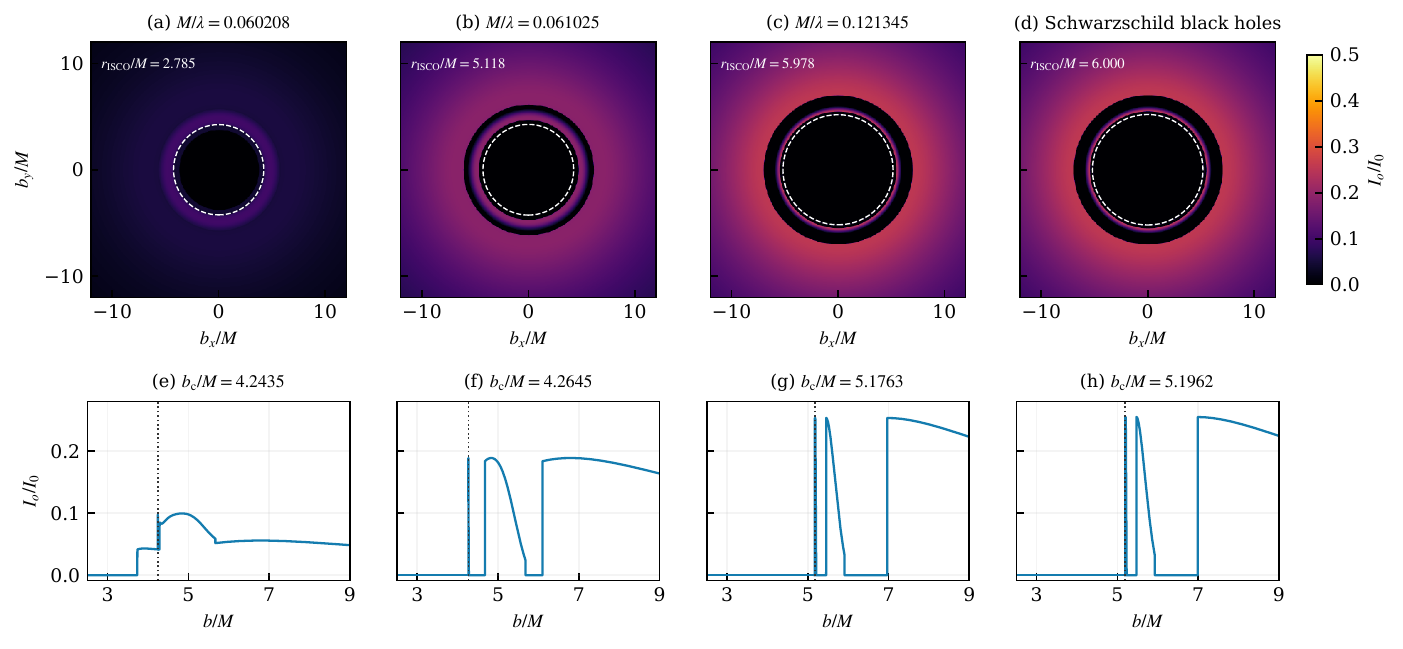}
 \caption{The upper row shows images of circular disks with inner edges at the ISCO.
 Panels (a), (b), and (c) show the $\kappa=100$ backgrounds at fixed $\lambda$ and $M/\lambda\simeq0.060208$, $0.061025$, and $0.121345$, respectively.
 Panels (a) and (b) lie on opposite sides of the transition from one to three MSCOs in Fig.~\ref{fig:isco_metric_transition}.
 Panel (d) shows the reference image for Schwarzschild black holes with $\risco=6M$, which is independent of mass in the displayed $b/M$ coordinates under the same emission model.
  All images use the same inner-edge intensity $I_0$, linear color scale $0\leq I_o/I_0\leq0.5$, radial emissivity index $p=2$, and outer disk radius $30M$.
 White dashed circles mark $b=\bcrit$.
 The lower panels (e)--(h) show the corresponding radial intensity profiles $I_o(b)/I_0$ as blue solid curves on a common vertical scale.
 Vertical dotted lines mark $\bcrit/M$.
 Panels (e) and (f) show that emission at $b<\bcrit$ disappears when the disk inner edge set by the ISCO moves outward between the backgrounds in panels (a) and (b), while $\bcrit/M$ changes only slightly.}
 \label{fig:isco_transition_images}
\end{figure}

To examine how the mass dependence of the ISCO affects the appearance of a circular disk, we present images along the radially stable $\kappa=100$ branch at fixed $\lambda$ in Fig.~\ref{fig:isco_transition_images}, together with a reference image for Schwarzschild black holes.
The masses $M/\lambda\simeq0.060208$ and $0.061025$ lie on opposite sides of the formation of the outer MSCO pair, while $M/\lambda\simeq0.121345$ shows the disk at higher mass.
Panels (a)--(d) display the images, and panels (e)--(h) give the corresponding radial intensity profiles, with a common inner-edge intensity $I_0$ for all four disks.

Comparing panels (a,b) and (e,f), we find that emission at $b<\bcrit$ disappears when the ISCO moves outward across the creation of the outer MSCO pair.
To explain the loss of interior emission, we compare $\risco$ with the limiting first-crossing radius $r_1(\bcrit^-)=\lim_{b\to\bcrit^-}r_1(b)$, approached from the captured side.
A captured backward ray reaches successive disk crossings at decreasing radii, so the first crossing determines whether that ray can encounter the emitting disk.
At $M/\lambda\simeq0.060208$, $\risco/M\simeq2.785<r_1(\bcrit^-)/M\simeq3.291$, allowing captured rays to intersect the emitting disk and produce the nonzero interior intensity in panel (e).
At $M/\lambda\simeq0.061025$, $\risco/M\simeq5.118>r_1(\bcrit^-)/M\simeq3.312$, preventing captured rays from intersecting the emitting disk and leaving no resolved interior emission in panel (f).
The disappearance of emission at $b<\bcrit$ therefore follows from the outward shift of the disk inner edge, while $\bcrit/M$ changes only from about $4.2435$ to $4.2645$.
With $I_0$ fixed at the disk inner edge, the larger $\risco/M$ in panel (b) also increases the prescribed $I_{\mathrm{em}}/I_0$ at a common $r/M$ within the emitting regions of both disks in panels (a) and (b).

At $M/\lambda\simeq0.121345$, we find that the image and radial profile in panels (c) and (g) approach the corresponding results for Schwarzschild black holes in panels (d) and (h).
The values $\risco/M\simeq5.978$ and $\bcrit/M\simeq5.1763$ are close to $6$ and $3\sqrt{3}$ for Schwarzschild black holes.
Both the higher-mass scalarized disk and the disk around Schwarzschild black holes have no resolved emission at $b<\bcrit$ in the adopted model.

\section{Conclusions and Outlook}\label{sec:conclusions}
We have investigated the shadows, photon trajectories, and radiative images of static, spherically symmetric black holes with purely nonlinear scalarization in scalar--Gauss--Bonnet gravity.
For the quartic exponential coupling, we compared the mass dependence along the $\kappa=100$ and $1000$ branches and the coupling dependence at matched ADM masses for $\kappa=100,200,\ldots,1000$, keeping $\lambda$ fixed.
Using Schwarzschild black holes as a common reference, we obtained geometric and imaging results that distinguish the effects of scalarization, emitter motion, and the disk inner edge.

We find smaller shadows for radially stable scalarized black holes than for Schwarzschild black holes of the same mass throughout the sampled branch intervals.
The largest reductions occur at the low-mass ends, where we obtain approximately $21.7\%$ for $\kappa=100$ and $19.3\%$ for $\kappa=1000$.
For both couplings, $\bcrit/M$ increases monotonically with mass toward $3\sqrt{3}$ while remaining below that value.
At matched ADM mass, we also find that the shadow radius increases with $\kappa$ across the sampled coupling grid.
These results show that the close agreement between images of the $\kappa=1000$ backgrounds and Schwarzschild black holes in the common mass interval is not representative of the entire stable branch.
At lower masses, we obtain appreciable shadow-size deviations even for $\kappa=1000$.

We find that a monotonic change in shadow size can coexist with nonmonotonic changes in the photon-orbit index $\chi$ and photon-ring width $W_{\rm ph}/M$ as mass or coupling varies.
At the minimum of $\chi$ along the stable $\kappa=100$ branch, we obtain a shadow radius only about $3.55\%$ below the value for Schwarzschild black holes, but an index about $25.9\%$ below unity.
The difference between these relative deviations shows that a modest shadow-size difference can accompany a substantially larger change in near-critical photon dynamics.
The smaller index implies slower separation of neighboring rays per unit orbital angle and weaker asymptotic compression of successive high-order images toward the critical curve.
We also find that the largest photon-ring width need not occur at the smallest $\chi$.
The width depends on the accumulated angle along complete photon paths, including propagation away from the photon sphere, whereas $\chi$ measures local photon-orbit instability.
Our results thus establish the complementary information provided by shadow size, photon-orbit instability, and photon-ring width.

For the radially unstable segments examined here, we obtain much smaller geometric deviations from Schwarzschild black holes.
The shadow-size deviation does not exceed approximately $0.34\%$, and the largest departure of $\chi$ from unity is about $1.6\%$.
We find that optical indicators close to those of Schwarzschild black holes coexist with radial instability of the background, so optical similarity alone does not establish dynamical stability.
The clearest geometric departures in our sample occur on the radially stable branches.

For spherical emission, we find that radial infall produces a darker center relative to the bright ring than static emission on the same background.
In both models, the ring lies near the critical curve and moves outward in units of $M$ as mass increases along the stable $\kappa=100$ sequence.
We obtain different mass dependences of the normalized peak intensity $I_{\mathrm{pk}}/(j_0M)$ for the two emitter motions.
The static-emission peak varies nonmonotonically and crosses the reference value for Schwarzschild black holes, whereas the infall peak decreases toward its reference value from above.
We therefore find that the approach of the shadow radius toward the value for Schwarzschild black holes does not determine how the brightness peak approaches its own reference.
The peak comparison uses intensities averaged over annuli of width $0.01M$, since the ideal pointwise intensity diverges logarithmically at the critical curve in the adopted optically thin models.

For face-on thin disks with a common inner edge at $6M$, we find that circular motion lowers the observed intensity relative to static emission on the same background, without changing the photon paths or emitting radial interval.
We obtain a stronger change when the disk inner edge is placed at the ISCO of the low-mass $\kappa=100$ backgrounds, where $\risco/M$ is close to $2.8$.
Captured backward rays can then intersect the emitting disk, producing emission inside the geometric shadow boundary at $b<\bcrit$.

Along the stable $\kappa=100$ branch, we further find that the creation of an outer pair of marginally stable circular orbits separates the inner and outer stable regions and changes the disk edge selected by our ISCO prescription.
Because the disk occupies the stable region connected to infinity, the selected edge moves outward when the intervening unstable interval appears.
For the two representative backgrounds on opposite sides of this transition, we obtain $\risco/M\simeq2.79$ and $5.12$.
We find that emission at $b<\bcrit$ disappears across these backgrounds even though $\bcrit/M$ changes only slightly.
The loss of interior emission is therefore caused by the change in the emitting region associated with circular-orbit stability.
The accompanying brightness changes also depend on our choice to fix the comoving intensity at each disk's own inner edge, which changes the prescribed emissivity in the shared outer region when that edge moves.

We conclude that purely nonlinear scalarization can produce distinct changes in shadow size, near-critical photon dynamics, and the orbital structure of emitting matter, despite the absence of a linear tachyonic onset of scalarization in this model.
Our imaging results show that the geometric critical curve, bright ring, and extent of the dark central region respond differently to geometry and emission.
Interpreting these image features together therefore requires both the metric and the emission prescription.
These conclusions apply to the sampled static spherical backgrounds, prescribed optically thin emission with negligible backreaction, and face-on disks.
%The stability classification concerns radial perturbations, and the present comparisons do not establish observational bounds on the coupling.

A natural extension is to determine how the region with multiple marginally stable circular orbits evolves with $\kappa$ and how matter populates the disconnected stable regions.
An accretion model that includes transport and emission in the plunging region would test whether the loss of interior emission found with a sharp ISCO boundary remains visible when the emitting region is determined dynamically.
Varying the disk inclination and including frequency-dependent emission and absorption would further test which features persist beyond the present bolometric, face-on models.
Extensions to rotating scalarized backgrounds and stability analyses beyond radial perturbations would broaden the class of physically relevant solutions.
Finally, combining more complete emission models with mass and distance information and an observational response would be necessary to assess whether the geometric and radiative differences identified here can be distinguished in black-hole images.

\section*{Acknowledgments}
The author Hao Yang thanks Li Li and Yiqian Chen for helpful discussions.
The author also acknowledges assistance from ChatGPT and Codex in developing the numerical codes, checking analytical derivations, and improving the language and presentation of the manuscript.
The author takes full responsibility for all scientific results and the entire content of this paper.

\section*{Data and Code Availability}
The numerical data and code supporting the results of this study are available from the authors upon reasonable request.
The available materials include the black-hole background solutions, optical and intensity-profile data, analysis and plotting scripts, and the associated parameter files and documentation.

\appendix
\section{Supplementary Analytical Relations}
\label{app:derivations}
We derive the photon-orbit and timelike-orbit relations used in Sections~\ref{subsec:chi_definition} and~\ref{subsec:thin_disk}, retaining the metric and coordinate conventions of Section~\ref{sec:backgrounds}.

\subsection{Photon-orbit instability and asymptotic ring scaling}
\label{app:photon_scaling}
Consider the nondegenerate maximum of $U_{\mathrm{opt}}=A/r^2$ at $r=\rph$ that controls capture.
At this radius, the circular-orbit condition gives
\begin{equation}
 U_{\mathrm{opt}}''(\rph)
 =\frac{\rph^2 A''(\rph)-2A(\rph)}{\rph^4}<0.
 \label{eq:app_optical_curvature}
\end{equation}
Expanding Eq.~\eqref{eq:null_coordinate_time} at $b=\bcrit$ and retaining the leading radial displacement gives
\begin{equation}
 \left(\frac{d\delta r}{dt}\right)^2
 =-\frac{A(\rph)B(\rph)\bcrit^2}{2}
 U_{\mathrm{opt}}''(\rph)(\delta r)^2
 +\mathcal O\bigl((\delta r)^3\bigr).
 \label{eq:app_local_null_motion}
\end{equation}
The growing solution therefore has the Lyapunov exponent in Eq.~\eqref{eq:photon_lyapunov_exponent}, in agreement with the geodesic stability analysis of Ref.~\citep{CardosoEtAl2009}.
Using $\bcrit^2=\rph^2/A(\rph)$ and $\Omegaph=1/\bcrit$ then gives Eq.~\eqref{eq:photon_orbit_index}.
The factor $B(\rph)$ enters the instability rate even though the photon-orbit condition itself depends only on $A$.

To obtain the logarithmic growth of the accumulated angle, take a scattered ray with $\epsilon=b/\bcrit-1>0$ and introduce
\begin{equation}
 x_{\mathrm{ph}}\equiv\frac{r-\rph}{\rph},\qquad
 \alpha_{\mathrm{ph}}\equiv
 -\frac{\bcrit^2\rph^2}{2}U_{\mathrm{opt}}''(\rph)
 =\frac{\chi^2}{B(\rph)}>0.
 \label{eq:app_near_orbit_variables}
\end{equation}
Here $x_{\mathrm{ph}}$ is the fractional radial displacement, and $\alpha_{\mathrm{ph}}$ is the dimensionless curvature coefficient of the capture barrier.
The radial factor in the geodesic equation becomes
\begin{equation}
 1-b^2U_{\mathrm{opt}}(r)
 =-2\epsilon+\alpha_{\mathrm{ph}}x_{\mathrm{ph}}^2
 +\mathcal O\bigl(\epsilon^2,\epsilon x_{\mathrm{ph}}^2,x_{\mathrm{ph}}^3\bigr).
 \label{eq:app_barrier_expansion}
\end{equation}
Thus the outer turning point satisfies $(r_t-\rph)/\rph\simeq\sqrt{2\epsilon/\alpha_{\mathrm{ph}}}$.
Within the same approximation, Eq.~\eqref{eq:ray_angular_integrand} gives
\begin{equation}
 \left|\frac{d\psi}{dx_{\mathrm{ph}}}\right|
 \simeq\frac{1}{\sqrt{B(\rph)}
 \sqrt{\alpha_{\mathrm{ph}}x_{\mathrm{ph}}^2-2\epsilon}}.
 \label{eq:app_singular_angular_integrand}
\end{equation}
Choose a dimensionless matching radius $\ell_{\mathrm{ph}}$ satisfying
$\sqrt{2\epsilon/\alpha_{\mathrm{ph}}}\ll\ell_{\mathrm{ph}}\ll1$.
It separates the near-orbit region from the rest of the path and has no physical significance.
Including both the incoming and outgoing segments, the quadratic approximation to the near-orbit angular contribution is
\begin{align}
 \Delta\psi_{\mathrm{near}}^{(2)}
 &=\frac{2}{\chi}\operatorname{arcosh}
 \left(\frac{\sqrt{\alpha_{\mathrm{ph}}}\ell_{\mathrm{ph}}}{\sqrt{2\epsilon}}\right)
 \nonumber\\
 &=-\frac{1}{\chi}\ln\epsilon
 +\frac{2}{\chi}\ln\left(\sqrt{2\alpha_{\mathrm{ph}}}\ell_{\mathrm{ph}}\right)
 +\mathcal O\left(\frac{\epsilon}{\alpha_{\mathrm{ph}}\ell_{\mathrm{ph}}^2}\right).
 \label{eq:app_near_angle_integral}
\end{align}
The superscript $(2)$ denotes the quadratic expansion of the radial factor.
Matching this contribution to the remaining path recovers~\citep{Bozza2002} Eq.~\eqref{eq:near_critical_angle}.
For a ray whose two endpoints are at infinity, the finite constant can be written as
\begin{equation}
 \begin{split}
 C=\lim_{\ell_{\mathrm{ph}}\to0^+}\Bigg[&
 2\int_{\rph(1+\ell_{\mathrm{ph}})}^{\infty}
 \frac{\bcrit\sqrt{A/B}}{r^2\sqrt{1-\bcrit^2U_{\mathrm{opt}}}}\,dr
 \\
 &+\frac{2}{\chi}\ln\left(\sqrt{2\alpha_{\mathrm{ph}}}\ell_{\mathrm{ph}}\right)
 \Bigg].
 \end{split}
 \label{eq:app_regular_angle_constant}
\end{equation}
The integral contains the opposite logarithmic dependence on $\ell_{\mathrm{ph}}$, so the limit is finite and independent of the arbitrary matching radius.
Finite source and observer radii outside the photon sphere change the regular contribution, but leave the coefficient $1/\chi$ unchanged when the endpoints remain away from the orbit.
For endpoints at infinity, the deflection angle is $\Delta\psi-\pi$, so subtracting $\pi$ changes only the finite constant.

The same logarithm determines the asymptotic widths discussed in Section~\ref{subsec:chi_definition}.
Consider a sequence of high-order scattered images of a fixed emitting annulus outside the photon sphere, with successive orders adding an angle $\pi$ as in the face-on disk geometry.
Let $b_n^{(1)}$ and $b_n^{(2)}$ denote the two impact-parameter boundaries of its $n$th image.
The finite path contributions at each source boundary are independent of $n$ in the critical limit.
Inverting the logarithmic relation therefore gives
\begin{equation}
 b_n^{(i)}-\bcrit
 =K_i e^{-\pi\chi n}[1+o(1)],\qquad i=1,2,
 \label{eq:app_image_boundary_scaling}
\end{equation}
where $K_i$ are constants with dimensions of length, fixed by the source boundaries and the observer geometry, and $o(1)$ now denotes terms vanishing as $n\to\infty$.
For distinct boundaries, the image width is consequently
\begin{equation}
 w_n\equiv\left|b_n^{(2)}-b_n^{(1)}\right|
 =|K_2-K_1|e^{-\pi\chi n}[1+o(1)].
 \label{eq:app_image_width_scaling}
\end{equation}
This yields~\citep{KocherlakotaEtAl2024} the half-orbit compression factor $e^{-\pi\chi}$ in Eq.~\eqref{eq:half_orbit_compression}.
If successive orders instead differ by a full orbit, the corresponding factor is $e^{-2\pi\chi}$.
These are asymptotic relations for geometric image intervals and do not specify their brightness or flux.

\subsection{Timelike circular orbits and disk inner edges}
\label{app:msco}
For a minimally coupled timelike test particle, let $E_{\mathrm t}=A\,dt/d\tau$ and $L_{\mathrm t}=r^2d\varphi/d\tau$ denote the conserved energy and angular momentum per unit rest mass, where $\tau$ is proper time.
The normalization of the four-velocity gives
\begin{equation}
 \frac{A}{B}\left(\frac{dr}{d\tau}\right)^2+V_{\mathrm t}(r;L_{\mathrm t})=E_{\mathrm t}^{\,2},\qquad
 V_{\mathrm t}(r;L_{\mathrm t})=A\left(1+\frac{L_{\mathrm t}^{\,2}}{r^2}\right).
 \label{eq:app_timelike_radial_motion}
\end{equation}
Here $V_{\mathrm t}$ is the timelike effective potential at fixed angular momentum.
A circular orbit requires $V_{\mathrm t}=E_{\mathrm t}^{\,2}$ and $\partial_rV_{\mathrm t}=0$.
Solving these conditions gives
\begin{equation}
 E_{\mathrm t,c}^{\,2}=\frac{2A^2}{2A-rA'},\qquad
 L_{\mathrm t,c}^{\,2}=\frac{r^3A'}{2A-rA'}.
 \label{eq:app_circular_energy_momentum}
\end{equation}
The subscript $c$ denotes values on the circular-orbit family.
In the static exterior, the rotating timelike orbits used for the disk require $A'>0$ and $2A-rA'>0$.
Their angular velocity and time component of the four-velocity satisfy
\begin{equation}
 \Omega_c^2=\left(\frac{A L_{\mathrm t,c}}{r^2 E_{\mathrm t,c}}\right)^2
 =\frac{A'}{2r},\qquad
 u_e^t=\frac{E_{\mathrm t,c}}{A}=\frac{1}{\sqrt{A-rA'/2}},
 \label{eq:app_circular_motion}
\end{equation}
recovering the kinematic factors in Eq.~\eqref{eq:circular_disk_redshift}.

Radial stability follows from the curvature of $V_{\mathrm t}$ while holding $L_{\mathrm t}$ fixed.
The angular momentum is set equal to $L_{\mathrm t,c}$ only after differentiating the potential.
Using the stability function $\mathcal S$ defined in Eq.~\eqref{eq:disk_msco_condition}, we obtain
\begin{equation}
 \left.\frac{\partial^2V_{\mathrm t}}{\partial r^2}\right|_{L_{\mathrm t}=L_{\mathrm t,c}}
 =\frac{2\mathcal S}{r(2A-rA')},\qquad
 \frac{dL_{\mathrm t,c}^{\,2}}{dr}
 =\frac{2r^2\mathcal S}{(2A-rA')^2}.
 \label{eq:app_timelike_stability_criteria}
\end{equation}
The second expression differentiates the circular-orbit family and shows~\citep{Song2021ISCO} why the angular-momentum and effective-potential criteria identify the same marginal radii.
To express stability as a dynamical response, let $\delta r_{\mathrm t}$ be a small radial displacement of the particle and define $\Omega_r$ as its radial oscillation frequency measured in $t$.
Linearizing the radial equation and using $dt/d\tau=E_{\mathrm t,c}/A$ gives
\begin{equation}
 \frac{d^2\delta r_{\mathrm t}}{dt^2}+\Omega_r^2\delta r_{\mathrm t}=0,\qquad
 \Omega_r^2
 =\frac{AB}{2E_{\mathrm t,c}^{\,2}}
 \left.\frac{\partial^2V_{\mathrm t}}{\partial r^2}\right|_{L_{\mathrm t}=L_{\mathrm t,c}}
 =\frac{B\mathcal S}{2Ar}.
 \label{eq:app_radial_epicyclic_frequency}
\end{equation}
All coefficients are evaluated at the unperturbed circular radius.
Positive $\Omega_r^2$ describes radial oscillations, while negative $\Omega_r^2$ describes exponential departure from the orbit.
Since $A,B>0$, the sign is fixed by $\mathcal S$.
Thus $A$ and its derivatives determine the marginal radii, whereas $B$ also affects the oscillation or growth rate.
For Schwarzschild black holes, these expressions reduce to
\begin{equation}
 \mathcal S_{\Sch}=\frac{2M(r-6M)}{r^4},\qquad
 \Omega_{r,\Sch}^2=\frac{M}{r^3}\left(1-\frac{6M}{r}\right),
 \label{eq:app_schwarzschild_timelike_check}
\end{equation}
recovering stable circular orbits for $r>6M$ and marginal stability at $6M$.

When several physical roots of $\mathcal S=0$ exist, their locations alone do not specify the disk inner edge.
One must also identify which intervals admit timelike circular orbits and have $\mathcal S>0$.
As discussed~\citep{DelgadoEtAl2022} for disconnected stable-orbit regions, the innermost stable orbit need not bound the stable region connected to infinity.
Our prescription in Section~\ref{subsec:thin_disk} uses the inner boundary of that outer region and leaves any disconnected inner stable region without emission.

The appearance of an additional unstable interval can be understood directly from a local expansion.
Let $(r_*,M_*)$ identify a degenerate MSCO along a smooth family at fixed $\kappa$ and $\lambda$, satisfying $\mathcal S=\partial_r\mathcal S=0$.
Near a generic double root,
\begin{equation}
 \mathcal S(r,M/\lambda)
 =a_{\mathrm{MS}}\frac{M-M_*}{\lambda}
 +c_{\mathrm{MS}}(r-r_*)^2+\cdots,\qquad
 a_{\mathrm{MS}}c_{\mathrm{MS}}\ne0,
 \label{eq:app_msco_double_root}
\end{equation}
where $a_{\mathrm{MS}}=\partial\mathcal S/\partial(M/\lambda)$ and $c_{\mathrm{MS}}=\tfrac12\partial_r^2\mathcal S$ are evaluated at the degenerate orbit.
The derivatives are taken at fixed areal radius and fixed mass, respectively, with the coupling parameters held constant.
The two nearby marginal radii, denoted by $r_{\mathrm{MS},\pm}$, therefore satisfy to leading order
\begin{equation}
 r_{\mathrm{MS},\pm}-r_*
 \simeq\pm\sqrt{-\frac{a_{\mathrm{MS}}}{c_{\mathrm{MS}}}
 \frac{M-M_*}{\lambda}},
 \label{eq:app_msco_root_splitting}
\end{equation}
on the side of the transition where the radicand is positive.
For $c_{\mathrm{MS}}>0$, the two roots enclose an interval with $\mathcal S<0$.
If this interval forms within the previously connected outer stable region, the adopted disk inner edge switches to the larger new root.
The selected ISCO can consequently change discontinuously even when the metric varies smoothly with mass.
The later merger of the inner and middle roots need not produce another jump, because the outer stable region can retain the same boundary.
These relations explain the changes of orbital structure in Fig.~\ref{fig:isco_metric_transition} without requiring a discontinuity or singularity of the metric.
The critical masses reported in Section~\ref{subsec:thin_disk} are obtained from the simultaneous conditions $\mathcal S=\partial_r\mathcal S=0$, rather than from the separation of the representative plotted masses.

\bibliographystyle{unsrtnat}
\bibliography{references}

\begin{thebibliography}{25}
\providecommand{\natexlab}[1]{#1}
\providecommand{\url}[1]{\texttt{#1}}
\expandafter\ifx\csname urlstyle\endcsname\relax
  \providecommand{\doi}[1]{doi: #1}\else
  \providecommand{\doi}{doi: \begingroup \urlstyle{rm}\Url}\fi

\bibitem[{Event Horizon Telescope Collaboration}(2019)]{EHT2019M87I}
{Event Horizon Telescope Collaboration}.
\newblock First {M87 Event Horizon Telescope} results. {I}. the shadow of the
  supermassive black hole.
\newblock \emph{The Astrophysical Journal Letters}, 875:\penalty0 L1, 2019.
\newblock \doi{10.3847/2041-8213/ab0ec7}.

\bibitem[{Event Horizon Telescope Collaboration}(2022)]{EHT2022SgrAI}
{Event Horizon Telescope Collaboration}.
\newblock First {Sagittarius A* Event Horizon Telescope} results. {I}. the
  shadow of the supermassive black hole in the center of the {Milky Way}.
\newblock \emph{The Astrophysical Journal Letters}, 930:\penalty0 L12, 2022.
\newblock \doi{10.3847/2041-8213/ac6674}.

\bibitem[{Event Horizon Telescope
  Collaboration}(2024{\natexlab{a}})]{EHT2024M87Persistent}
{Event Horizon Telescope Collaboration}.
\newblock The persistent shadow of the supermassive black hole of {M 87}. {I}.
  observations, calibration, imaging, and analysis.
\newblock \emph{Astronomy \& Astrophysics}, 681:\penalty0 A79,
  2024{\natexlab{a}}.
\newblock \doi{10.1051/0004-6361/202347932}.

\bibitem[{Event Horizon Telescope
  Collaboration}(2024{\natexlab{b}})]{EHT2024SgrAPolarization}
{Event Horizon Telescope Collaboration}.
\newblock First {Sagittarius A* Event Horizon Telescope} results. {VII}.
  polarization of the ring.
\newblock \emph{The Astrophysical Journal Letters}, 964:\penalty0 L25,
  2024{\natexlab{b}}.
\newblock \doi{10.3847/2041-8213/ad2df0}.

\bibitem[{Event Horizon Telescope Collaboration}(2025)]{EHT2025M87Variability}
{Event Horizon Telescope Collaboration}.
\newblock Horizon-scale variability of {M87*} from 2017--2021 {EHT}
  observations.
\newblock \emph{Astronomy \& Astrophysics}, 704:\penalty0 A91, 2025.
\newblock \doi{10.1051/0004-6361/202555855}.

\bibitem[Perlick and Tsupko(2022)]{PerlickTsupko2022}
Volker Perlick and Oleg~Yu. Tsupko.
\newblock Calculating black hole shadows: Review of analytical studies.
\newblock \emph{Physics Reports}, 947:\penalty0 1--39, 2022.
\newblock \doi{10.1016/j.physrep.2021.10.004}.

\bibitem[Gralla et~al.(2019)Gralla, Holz, and Wald]{GrallaHolzWald2019}
Samuel~E. Gralla, Daniel~E. Holz, and Robert~M. Wald.
\newblock Black hole shadows, photon rings, and lensing rings.
\newblock \emph{Physical Review D}, 100:\penalty0 024018, 2019.
\newblock \doi{10.1103/PhysRevD.100.024018}.

\bibitem[Antoniou et~al.(2018)Antoniou, Bakopoulos, and
  Kanti]{AntoniouEtAl2018}
G.~Antoniou, A.~Bakopoulos, and P.~Kanti.
\newblock Evasion of no-hair theorems and novel black-hole solutions in
  {Gauss-Bonnet} theories.
\newblock \emph{Physical Review Letters}, 120:\penalty0 131102, 2018.
\newblock \doi{10.1103/PhysRevLett.120.131102}.

\bibitem[Doneva and Yazadjiev(2018)]{Doneva:2017bvd}
Daniela~D. Doneva and Stoytcho~S. Yazadjiev.
\newblock {New Gauss-Bonnet Black Holes with Curvature-Induced Scalarization in
  Extended Scalar-Tensor Theories}.
\newblock \emph{Phys. Rev. Lett.}, 120\penalty0 (13):\penalty0 131103, 2018.
\newblock \doi{10.1103/PhysRevLett.120.131103}.

\bibitem[Silva et~al.(2018)Silva, Sakstein, Gualtieri, Sotiriou, and
  Berti]{SilvaEtAl2018}
Hector~O. Silva, Jeremy Sakstein, Leonardo Gualtieri, Thomas~P. Sotiriou, and
  Emanuele Berti.
\newblock Spontaneous scalarization of black holes and compact stars from a
  {Gauss-Bonnet} coupling.
\newblock \emph{Physical Review Letters}, 120:\penalty0 131104, 2018.
\newblock \doi{10.1103/PhysRevLett.120.131104}.

\bibitem[Doneva et~al.(2024)Doneva, Ramazano{\u g}lu, Silva, Sotiriou, and
  Yazadjiev]{DonevaEtAl2024Review}
Daniela~D. Doneva, Fethi~M. Ramazano{\u g}lu, Hector~O. Silva, Thomas~P.
  Sotiriou, and Stoytcho~S. Yazadjiev.
\newblock Spontaneous scalarization.
\newblock \emph{Reviews of Modern Physics}, 96:\penalty0 015004, 2024.
\newblock \doi{10.1103/RevModPhys.96.015004}.

\bibitem[Doneva and Yazadjiev(2022)]{DonevaYazadjiev2022}
Daniela~D. Doneva and Stoytcho~S. Yazadjiev.
\newblock Beyond the spontaneous scalarization: New fully nonlinear mechanism
  for the formation of scalarized black holes and its dynamical development.
\newblock \emph{Physical Review D}, 105\penalty0 (4):\penalty0 L041502, 2022.
\newblock \doi{10.1103/PhysRevD.105.L041502}.

\bibitem[Bl{\'a}zquez-Salcedo et~al.(2022)Bl{\'a}zquez-Salcedo, Doneva, Kunz,
  and Yazadjiev]{BlazquezSalcedoEtAl2022}
Jose~Luis Bl{\'a}zquez-Salcedo, Daniela~D. Doneva, Jutta Kunz, and Stoytcho~S.
  Yazadjiev.
\newblock Radial perturbations of scalar-{Gauss-Bonnet} black holes beyond
  spontaneous scalarization.
\newblock \emph{Physical Review D}, 105:\penalty0 124005, 2022.
\newblock \doi{10.1103/PhysRevD.105.124005}.

\bibitem[Cunha et~al.(2019)Cunha, Herdeiro, and Radu]{CunhaHerdeiroRadu2019}
Pedro V.~P. Cunha, Carlos A.~R. Herdeiro, and Eugen Radu.
\newblock Spontaneously scalarized {Kerr} black holes in extended
  scalar-tensor--{Gauss-Bonnet} gravity.
\newblock \emph{Physical Review Letters}, 123:\penalty0 011101, 2019.
\newblock \doi{10.1103/PhysRevLett.123.011101}.

\bibitem[Fernandes et~al.(2024)Fernandes, Burrage, Eichhorn, and
  Sotiriou]{FernandesEtAl2024}
Pedro G.~S. Fernandes, Clare Burrage, Astrid Eichhorn, and Thomas~P. Sotiriou.
\newblock Shadows and properties of spin-induced scalarized black holes with
  and without a {Ricci} coupling.
\newblock \emph{Physical Review D}, 109:\penalty0 104033, 2024.
\newblock \doi{10.1103/PhysRevD.109.104033}.

\bibitem[Wang et~al.(2024)Wang, Meng, Kuang, and Liao]{WangEtAl2024}
Xi-Jing Wang, Yuan Meng, Xiao-Mei Kuang, and Kai Liao.
\newblock Distinguishing black holes with and without spontaneous scalarization
  in {Einstein-scalar-Gauss-Bonnet} theories via optical features.
\newblock \emph{The European Physical Journal C}, 84:\penalty0 1243, 2024.
\newblock \doi{10.1140/epjc/s10052-024-13612-w}.

\bibitem[Cardoso et~al.(2009)Cardoso, Miranda, Berti, Witek, and
  Zanchin]{CardosoEtAl2009}
Vitor Cardoso, Alex~S. Miranda, Emanuele Berti, Helvi Witek, and Vilson~T.
  Zanchin.
\newblock Geodesic stability, {Lyapunov} exponents and quasinormal modes.
\newblock \emph{Physical Review D}, 79:\penalty0 064016, 2009.
\newblock \doi{10.1103/PhysRevD.79.064016}.

\bibitem[Kocherlakota et~al.(2024)Kocherlakota, Rezzolla, Roy, and
  Wielgus]{KocherlakotaEtAl2024}
Prashant Kocherlakota, Luciano Rezzolla, Rittick Roy, and Maciek Wielgus.
\newblock Prospects for future experimental tests of gravity with black hole
  imaging: Spherical symmetry.
\newblock \emph{Physical Review D}, 109:\penalty0 064064, 2024.
\newblock \doi{10.1103/PhysRevD.109.064064}.

\bibitem[Johnson et~al.(2020)Johnson, Lupsasca, Strominger, Wong, Hadar, Kapec,
  Narayan, Chael, Gammie, Galison, Palumbo, Doeleman, Blackburn, Wielgus,
  Pesce, Farah, and Moran]{JohnsonEtAl2020}
Michael~D. Johnson, Alexandru Lupsasca, Andrew Strominger, George~N. Wong,
  Shahar Hadar, Daniel Kapec, Ramesh Narayan, Andrew Chael, Charles~F. Gammie,
  Peter Galison, Daniel C.~M. Palumbo, Sheperd~S. Doeleman, Lindy Blackburn,
  Maciek Wielgus, Dominic~W. Pesce, Joseph~R. Farah, and James~M. Moran.
\newblock Universal interferometric signatures of a black hole's photon ring.
\newblock \emph{Science Advances}, 6:\penalty0 eaaz1310, 2020.
\newblock \doi{10.1126/sciadv.aaz1310}.

\bibitem[Narayan et~al.(2019)Narayan, Johnson, and
  Gammie]{NarayanJohnsonGammie2019}
Ramesh Narayan, Michael~D. Johnson, and Charles~F. Gammie.
\newblock The shadow of a spherically accreting black hole.
\newblock \emph{The Astrophysical Journal Letters}, 2019.
\newblock \doi{10.3847/2041-8213/ab518c}.

\bibitem[Kocherlakota and Rezzolla(2022)]{KocherlakotaRezzolla2022}
Prashant Kocherlakota and Luciano Rezzolla.
\newblock Distinguishing gravitational and emission physics in black hole
  imaging: spherical symmetry.
\newblock \emph{Monthly Notices of the Royal Astronomical Society},
  513\penalty0 (1):\penalty0 1229--1243, 2022.
\newblock \doi{10.1093/mnras/stac891}.

\bibitem[Delgado et~al.(2022)Delgado, Herdeiro, and Radu]{DelgadoEtAl2022}
Jorge F.~M. Delgado, Carlos A.~R. Herdeiro, and Eugen Radu.
\newblock Equatorial timelike circular orbits around generic ultracompact
  objects.
\newblock \emph{Physical Review D}, 105:\penalty0 064026, 2022.
\newblock \doi{10.1103/PhysRevD.105.064026}.

\bibitem[Bozza(2002)]{Bozza2002}
Valerio Bozza.
\newblock Gravitational lensing in the strong field limit.
\newblock \emph{Physical Review D}, 66:\penalty0 103001, 2002.
\newblock \doi{10.1103/PhysRevD.66.103001}.

\bibitem[Song(2021)]{Song2021ISCO}
Yong Song.
\newblock The evolutions of the innermost stable circular orbits in dynamical
  spacetimes.
\newblock \emph{European Physical Journal C}, 81:\penalty0 875, 2021.
\newblock \doi{10.1140/epjc/s10052-021-09623-6}.

\bibitem[Deliyski et~al.(2026)Deliyski, Gyulchev, Doneva, Nedkova, and
  Yazadjiev]{DeliyskiEtAl2026}
Valentin~O. Deliyski, Galin~N. Gyulchev, Daniela~D. Doneva, Petya~G. Nedkova,
  and Stoytcho~S. Yazadjiev.
\newblock Polarized equatorial emission around {Kerr} black holes with
  synchronized scalar hair. {I}. direct images.
\newblock \emph{European Physical Journal C}, 86:\penalty0 908, 2026.
\newblock \doi{10.1140/epjc/s10052-026-16103-2}.

\end{thebibliography}
\end{document}